\documentclass[10pt,twocolumn,journal]{IEEEtran}
\usepackage[T1]{fontenc}
\usepackage{epsfig,psfrag}
\usepackage[latin9]{inputenc}
\usepackage{amsmath,amsfonts,amssymb,amsxtra,bm}
\usepackage{graphicx}
\usepackage{tikz}
\usepackage{color}
\usepackage{cite}
\usepackage{setspace}
\usepackage{tikz}
\usepackage{pgfplots}
\usepackage{caption}
\usepackage{notation}
\usepackage{dblfloatfix}
\usepackage{cuted}
\usepackage{lipsum}
\usepackage{mathtools}
\usepackage{float}

\newcommand{\ist}{\hspace*{.3mm}}
\newcommand{\rmv}{\hspace*{-.3mm}}

\newcommand{\be}{\begin{equation}}
\newcommand{\ee}{\end{equation}}
\newcommand{\iist}{\hspace*{1mm}}
\newcommand{\rrmv}{\hspace*{-1mm}}

\newcommand{\nn}{\nonumber}

\newcommand{\T}{\text{T}}

\DeclareMathAlphabet{\mathpzc}{OT1}{pzc}{m}{it}

\newcommand{\va}[1]{#1}

\DeclareMathAlphabet{\mathpzc}{OT1}{pzc}{m}{it}

\newcommand{\RFSst}{\Set{X}}
\newcommand{\RFSstR}{\RS{X}}	
\newcommand{\st}{\V{x}}
\newcommand{\stR}{\RV{x}}
\newcommand{\ex}{r}
\newcommand{\sd}{f}

\newcommand{\me}{z}
\newcommand{\meR}{\rv{z}}
\newcommand{\mev}{\V{z}}
\newcommand{\mevR}{\RV{z}}

\newcommand{\ass}{\va{a}}								
\newcommand{\assv}{\V{a}}								
\newcommand{\assR}{\rv{a}}							
\newcommand{\assvR}{\RV{a}}

\newcommand{\assw}{\beta}
\newcommand{\su}{p_{\text{S}}}

\newcommand{\PHD}{\lambda}
\newcommand{\card}{\rho}

\allowdisplaybreaks
\pgfplotsset{compat=1.18}
\begin{document}

\title{Association-Based Track-Before-Detect \\with Object Contribution Probabilities}

\author{Thomas Kropfreiter, Jason L. Williams, and Florian Meyer

\thanks{The material presented in this work was supported by the Austrian Science Fund (FWF) under Grant J\,4726-N and the National Science Foundation (NSF) under CAREER Award No. 2146261.}

\thanks{T.~Kropfreiter  is with the Institute of Telecommunications, TU Wien, Vienna, Austria (e-mail: thomas.kropfreiter@tuwien.ac.at). The material presented in this paper was performed as a postdoctoral researcher at the Scripps Institution of Oceanography, University of California San Diego, La Jolla, CA, USA}

\thanks{J.~L.~Williams is with Whipbird Signals, Brisbane, Australia (jason@whipbird.au).}

\thanks{F.~Meyer is with the Scripps Institution of Oceanography and the Department of Electrical and Computer Engineering, University of California San Diego, La Jolla, CA, USA (e-mail: flmeyer@ucsd.edu).}\vspace{-5mm}}

\maketitle

\begin{abstract}

Multiobject tracking provides situational awareness that enables new applications for modern convenience, applied ocean sciences, public safety, and homeland security. 
In many multiobject tracking applications, including radar and sonar tracking, after coherent prefiltering of the received signal, measurement data is typically structured in cells, where each cell represent, e.g., a different range and bearing value. 
While conventional \textit{detect-then-track} (DTT) multiobject tracking approaches convert the cell-structured data within a detection phase into so-called point measurements in order to reduce the amount of data, \textit{track-before-detect} (TBD) methods process the cell-structured data directly, avoiding a potential information loss.
However, many TBD tracking methods are computationally intensive and achieve a reduced tracking accuracy when objects interact, i.e., when they come into close proximity.
We here counteract these difficulties by introducing the concept of probabilistic object-to-cell contributions.
As many conventional DTT methods, our approach uses a probabilistic association of objects with data cells, and a new object contribution model with corresponding object contribution probabilities to further associate cell contributions to objects that occupy the same data cell. 
Furthermore, to keep the computational complexity and filter runtimes low, we here use an efficient Poisson multi-Bernoulli filtering approach in combination with the application of belief propagation for fast probabilistic data association.
We demonstrate numerically that our method achieves significantly increased tracking performance compared to state-of-the-art TBD tracking approaches, where performance differences are particularly pronounced when multiple objects interact.
\vspace{1mm}
\end{abstract}

\begin{IEEEkeywords}
Multiobject tracking, 
multitarget tracking,
track-before-detect, 
random finite 
\vspace{-3mm}
sets.
\end{IEEEkeywords}

\section{Introduction}
\label{sec:int}
Multiobject tracking aims to estimate the time-dependent states of a time-dependent number of objects from data provided by one or several sensors. 
The raw data provided by the sensor(s) is preprocessed in multiple stages. First, a bank of matched filters in time and space is employed to maximize the signal-to-noise ratio (SNR). As a result, data cells, e.g., in range-bearing domain, are obtained. In conventional  \emph{detect-then-track} (DTT) multiobject tracking \cite{Bar11,Cha11,Koc14,Bar95,For83,Rei79,Mah07,Mah03,Mah07b,Vo13,Wil:J15,Mey18Proc,Mey17MSBP,Kro16,Kro24Coa} few so-called point measurements are extracted from data cells to reduce data rate. 
The reduced data rate results in a reduced computational complexity of the corresponding multiobject tracking method. 
More precisely, point measurements are obtained by applying a threshold test or detector to the data cells, which breaks up the cell structure and thereby converts cell indexes into, e.g., range and bearing measurements. 
In the detection phase, however, certain object-related information is inevitably lost.

\emph{Track-before-detect} (TBD) \cite{Ton98,Moy11,Bar85TBD,Dav18TBD,Ris13,Ris19,Vo10TBD,Kim21,Sal01TBD,Ort02TBD,Boe03TBD,Ris20Ber,Lia23TBD,Dav24TBD} directly uses data cells as measurements for multiobject tracking and thus avoids a potential loss of information in a detection phase. This is expected to lead to improved tracking performance, especially in challenging scenarios with low SNR\vspace{-3mm}. 

\subsection{State of the Art}
\vspace{-.5mm}

Irrespective of the measurement model that is in force, the multiobject state can either be modeled by a random vector \cite{Bar11,Cha11,Koc14,Bar95,For83,Rei79,Mey18Proc,Mey17MSBP,Kro24Coa} or a random set, more precisely, a random finite set (RFS) \cite{Mah07,Mah03,Mah07b,Vo13,Wil:J15,Kro16}.
Existing RFS-based DTT tracking methods include the probability hypothesis density (PHD) filter \cite{Mah03,Vo05}, the cardinalized PHD filter \cite{Mah07b}, multi-Bernoulli (MB) filters \cite{Vo09,Gar19MBM}, hybrid Poisson/MB filters \cite{Wil:J15,Kro16,Gar18PMBM,Wil12}, labeled RFS filters \cite{Vo13,Vo14,Reu14,Kro19LMB}, and filters based on RFSs of trajectories \cite{Gar20SoT,Gar19SoTPHD}.

Existing TBD methods can furthermore be distinguished between batch processing methods and sequential Bayesian estimation methods. 
Batch processing methods include approaches that are based on maximum likelihood estimation \cite{Ton98}, the Hough transform \cite{Moy11}, and dynamic programming techniques \cite{Bar85TBD}. 
However, due to their high computational complexity, they are often unsuitable for real-time operation. 
A well-established exception for real-time tracking is the histogram probabilistic multi-hypothesis Tracker (H-PMHT) \cite{Dav18TBD}, which is based on the expectation-maximization algorithm. 
However, tuning the parameters of the H-PMHT is known to be difficult \cite{Kim21}. 

Early sequential Bayesian vector-based TBD tracking methods rely on particle filtering \cite{Sal01TBD,Ort02TBD,Boe03TBD}.
While the method in \cite{Sal01TBD} is designed for the tracking of a single object, the methods in \cite{Ort02TBD,Boe03TBD} are limited to a known and small number of objects.
On the other hand, RFS-based methods are inherently capable of tracking an unknown number of objects.
These methods include the (single) Bernoulli filter \cite{Ris13,Ris20Ber} for single-object tracking and MB filters \cite{Vo10TBD,Kim21,Dav24TBD} for multiobject tracking.
However, the MB filter in \cite{Vo10TBD} performs poorly when objects interact, i.e., come close to each other, and all three MB filters rely on heuristics for generating new Bernoulli components.

Of particular interest here is the track-oriented marginal multi-Bernoulli/Poisson filter \cite{Wil:J15,Mey18Proc,Kro16,Gar18PMBM,Kro21TBD}, which will be simply referred to as the Poisson/MB (PMB) filter in the following. The PMB filter was both proposed for DTT multiobject tracking \cite{Wil:J15,Mey18Proc,Kro16,Gar18PMBM} and recently also for TBD multiobejct tracking \cite{Kro21TBD}.
The PMB filter models the multiobject state as the union of a Poisson RFS and an MB RFS.
Whereas the MB part is used for the tracking of already detected objects, the Poisson part models objects that are still undetected. 
The modeling of undetected objects facilitates the reliable generation of new Bernoulli components.
The modified version of the PMB filter in \cite{Wil12} extends the use of the Poisson RFS from the modeling of undetected objects to the tracking of objects that are unlikely to exist by introducing the concept of recycling.
More precisely, in the recycling step, Bernoulli components with a low existence probability are transferred from the MB part of the multiobject state to the Poisson part of the multiobject state instead of pruning them.  
The computational complexity of the PMB filter scales exponentially due to the high-dimensional marginalization of an inherent object-measurement association probability mass function (pmf). 
However, a fast implementation with only linear complexity scaling can be obtained by an approximate belief propagation (BP) based marginalization \cite{Mey18Proc,Wil14}. 
The PMB filter has also been derived and extended to multiple sensors using the framework of factor graphs and the sum-product algorithm \cite{Mey18Proc,Mey17MSBP}. It has been demonstrated that it can outperform existing DTT methods and that it is highly scalable in all relevant system parameters \cite{Wil:J15,Mey18Proc,Mey17MSBP,Kro16}\vspace{-3mm}.

\subsection{Contribution}
\label{sec:contr}

In this paper, we introduce a PMB filter for TBD that relies on a new measurement model based on the notion of random object contributions and corresponding object contribution probabilities. 
More precisely, we assume that each object contributes to at most a single data cell, which is in contrast to other TBD methods that are based on superpositional object contrition models \cite{Lia23TBD,Dav24TBD}. 
Within the filter derivation, we apply certain approximations, including efficient BP-based marginalization operations, leading to our PMB filter that exhibits a linear complexity scaling.
We verify the performance of our proposed method numerically and compare it to state-of-the-art TBD methods \cite{Vo10TBD,Lia23TBD,Dav24TBD}. 
An important result of our work is that our proposed method using association-based cell contributions achieves very good performance even when the sensor data was generated according to a superimposed data cell model. 
This implies that the approximations made by methods with superpositional models have a stronger impact on the overall tracking performance than the moderate model mismatch of our proposed method.

The main contributions of this paper can be summarized as \vspace{-.3mm} follows:
\begin{itemize}
\item We propose a novel TBD measurement model based on random object-to-cell contributions with corresponding object contribution probabilities. 
\vspace{.5mm}
\item We derive a new PMB filter for TBD that uses BP for efficient probabilistic object-to-cell associations.
\vspace{.5mm}
\item We demonstrate performance improvements of our TBD PMB filter compared to state-of-the-art TBD methods.

\end{itemize}
This paper differs from \cite{Kro21TBD} in that we extend the TBD measurement model by a probabilistic object contribution model with corresponding object contribution probabilities.
This leads to a new TBD PMB filter that achieves improved tracking accuracy 
in scenarios with multiple interacting objects, i.e., when multiple objects come in close proximity and contribute to the same resolution cell\vspace{-3mm}.

\subsection{Paper Organization and Notation}
\vspace{-.5mm}

This paper is structured as follows. 
Section \ref{sec:fund} reviews some fundamentals of RFSs.
Section \ref{sec:sys} presents the object state model together with the state transition model underlying our proposed PMB filter, and Section \ref{sec:measmod} presents the measurement model underlying the our PMB filter.
Section \ref{sec:pred} presents the prediction step of our PMB filter, and Sections \ref{sec:upd_step} and \ref{sec:up_st_two} present the corresponding update step.
Finally, in Section \ref{sec:num}, we provide numerical results underpinning the good performance of our proposed PMB filter compared to three state-of-the-art TBD filters.

We will use the following basic notation. Random variables are displayed in sans serif, upright fonts; their realizations in serif, italic fonts. Vectors and matrices are denoted by bold lowercase and uppercase letters, respectively. 
For example, $\rv{x}$ is a random variable and $x$ is its realization, and $\RV{x}$ is a random vector and $\V{x}$ is its realization. 
Random sets and their realizations are denoted by upright sans serif and calligraphic fonts, respectively. 
For example, $\RS{X}$ is a random set and $\Set{X}$ is its realization. 
We denote probability density functions (pdfs) by $f(\cdot)$ and pmfs by $p(\cdot)$. 
Further, $\Set{N}(\V{x}; \V{\mu},\BM{\Sigma})$ denotes the Gaussian pdf  (of random vector $\RV{x}$) with mean $\V{\mu}$ and covariance \vspace{0mm} matrix $\BM{\Sigma}$, $\Set{R}(x; \sigma)$ denotes the Rayleigh pdf (of scalar random variable $\rv{x}$) with scale parameter $\sigma$, and $\Set{P}(n;\mu)$ denotes the Poisson pmf (of scalar random integer $\rv{n}$) with mean parameter $\mu$.
The probability of an event is denoted by $\mathrm{Pr}\{\cdot\}$. The symbol $\propto$ indicates equality up to a normalization \vspace{-2mm} factor.

\section{RFS Fundamentals}
\vspace{1mm}
\label{sec:fund}

An RFS $\RFSstR \rmv=\rmv \{\stR^{(1)},\ldots,\stR^{({\sf n})}\}$ is a random variable where both the elements $\stR^{(i)}$ and the number of elements ${\sf n} \rmv= \rmv|\RFSstR|$, i.e., the cardinality of $\RFSstR$, are random quantities.
Thus, an RFS consists of a random number of random vectors.
The statistics of an RFS $\RFSstR$ can be described by its multiobject pdf $f(\RFSst)$ \cite{Mah07}. 
For any realization $\RFSst \rmv=\rmv \big\{\st^{(1)}\rmv,\ldots,\st^{(n)}\big\}$, the multiobject pdf $f(\RFSst)$ is given by
\vspace{0mm} 
\begin{equation}
\label{eq:fund0}
f(\RFSst) \ist=\ist n! \ist \card(n) \ist f_n(\st^{(1)}\rmv,\ldots,\st^{(n)})
\vspace{1mm}
\end{equation}
where $\card(n) \rmv\triangleq\rmv \mathrm{Pr}\{|\RFSstR| \rmv=\rmv n\}$, $n \rmv\in \mathbb{N}_0$ is the pmf of the cardinality ${\sf n} \rmv=\rmv |\RFSstR|$, and $f_n(\st^{(1)}\rmv,\ldots,\st^{(n)})$ is the joint pdf of the random vectors $\stR^{(1)}\rmv,\ldots,\stR^{(n)}$ that is permutation invariant with respect to its arguments $\st^{(i)}$\rmv.
While conventional Riemann integration cannot be applied to functions with set arguments, one can define a set integral 
of a real-valued set function $g(\RFSst)$ according to \cite{Mah07}
\begin{align}
&\int g(\RFSst) \ist\ist \delta \RFSst \nn\\
&\hspace{3mm}\triangleq \sum_{n\ist=\ist 0}^{\infty} \ist \frac{1}{n!} \ist \int_{\mathbb{R}^{n n_x}} \! g(\{\st^{(1)}\rmv,\ldots,\st^{(n)}\}) 
  \ist\ist \text{d}\st^{(1)}\rmv\cdots \text{d}\st^{(n)}. \label{eq:RFSfund_SetInt} \\[-2mm]
\nn\\[-7.5mm]
\nn
\end{align}
Note that for each cardinality value $n \rmv=\rmv |\RFSst|$ there is exactly one summation term. The multiobject pdf $f(\RFSst)$ integrates to one with respect to the set integral \eqref{eq:RFSfund_SetInt}.
In the following, we will review six important types of RFSs \cite{Mah07,Wil:J15,Mey18Proc} that are relevant for the underlying work.

A \textit{Bernoulli RFS} $\RFSstR$ is characterized by an existence probability $\ex$ and a spatial pdf $\sd(\st)$. 
The Bernoulli RFS is empty with probability $1-\ex$ and it contains exactly one element $\stR$ with probability $\ex$. In the latter case, $\stR$ is distributed according to $\sd(\st)$.
Hence, the multiobject pdf of the Bernoulli RFS is given\vspace{-1mm} by 
\begin{equation}
\label{eq:fund2}
f^{\text{B}}(\RFSst) \ist=
\begin{cases} 
1 \!-\rmv \ex , 			 &\RFSst \!=\rmv \emptyset, \\[-.6mm]
\ex \ist \sd(\st) \ist, &\RFSst \!=\! \{\st\}, \\[-.6mm]
0, & \text{otherwise}.
\end{cases}
\vspace{0mm}
\end{equation}

A \textit{multi-Bernoulli (MB) RFS} $\RFSstR$ is the union of a fixed number $J$ of statistically independent Bernoulli RFSs $\RFSstR^{(j)}\rmv$, $j \in \Set{J} \triangleq \{1,\ldots,J\}$.
Each Bernoulli RFS is thereby described by a Bernoulli pdf $f^{(j)}(\RFSst^{(j)})$, which in turn is characterized by a corresponding 
existence probability $\ex^{(j)}$ and a spatial pdf $\sd^{(j)}(\st)$ (cf. Eq.\! \eqref{eq:fund2}). 
An expression for the multiobject pdf $f^{\text{MB}}(\RFSst)$ can be obtained by sequentially applying the set convolution operation \cite{Mah07} to all the individual Bernoulli pdfs $f^{(j)}(\RFSst^{(j)})$. For $n = |\RFSst| \leq J$, this leads to 
\begin{equation}
\label{eq:fund3}
f^{\text{MB}}(\RFSst) = \sum_{\RFSst^{(1)}\uplus\ldots\uplus\ist\RFSst^{(J)} =\ist \RFSst} \prod_{j\ist\in\ist \Set{J}}  \ist\ist f^{(j)}(\RFSst^{(j)}) \ist.
\end{equation}
Here, 
\vspace{0.3mm}
$\sum_{\RFSst^{(1)}\uplus\ldots\uplus\RFSst^{(J)} = \RFSst}$ denotes the sum over all disjoint decompositions of $\RFSst$ into sets $\RFSst^{(j)}$\rmv, $j\rmv\in\rmv\Set{J}$ such that $\RFSst^{(1)}\cup\ldots\cup\RFSst^{(J)} = \RFSst$. 
For example, for $J \rmv=\rmv 3$, the multiobject pdf $f^{\text{MB}}(\RFSst)$ evaluated for $\RFSst = \big\{\st^{(1)}\rmv\rmv,\st^{(2)}\big\}$ is given as
\begin{align*}
&f^{\text{MB}}(\{\st^{(1)}\rmv,\st^{(2)}\}) \nn \\ 
&\hspace{10mm}=\ist \ex^{(1)} \ist \sd^{(1)}(\st^{(1)}) \ist \ex^{(2)} \ist \sd^{(2)}(\st^{(2)}) \ist (1\!-\rmv \ex^{(3)}) \nn \\
&\hspace{10mm}+\ist\ex^{(1)} \ist\sd^{(1)}(\st^{(2)}) \ist\ex^{(2)} \ist\sd^{(2)}(\st^{(1)}) \ist(1\!-\rmv \ex^{(3)}) \nn \\
&\hspace{10mm}+\ist\ex^{(1)} \ist\sd^{(1)}(\st^{(1)}) \ist(1\!-\rmv \ex^{(2)}) \ist\ex^{(3)} \ist\sd^{(3)}(\st^{(2)}) \nn \\
&\hspace{10mm}+\ist\ex^{(1)} \ist\sd^{(1)}(\st^{(2)}) \ist(1\!-\rmv \ex^{(2)}) \ist\ex^{(3)} \ist\sd^{(3)}(\st^{(1)}) \nn \\
&\hspace{10mm}+ \ist(1\!-\rmv \ex^{(1)}) \ist\ex^{(2)} \ist\sd^{(2)}(\st^{(1)}) \ist\ex^{(3)} \ist\sd^{(3)}(\st^{(2)}) \nn \\
&\hspace{10mm}+ \ist(1\!-\rmv \ex^{(1)}) \ist\ex^{(2)} \ist\sd^{(2)}(\st^{(2)}) \ist\ex^{(3)} \ist\sd^{(3)}(\st^{(1)}) \ist. 
\end{align*}
For $n \rmv>\rmv J$, we have $f^{\text{MB}}(\RFSst) \rmv=\rmv 0$.

An \textit{MB mixture (MBM) RFS} $\RFSstR$ is a weighted sum of MB RFSs. 
Without loss of generality, it is assumed that all MB RFSs have the same number of Bernoulli components $J$. 
For any realization $\RFSst = \{\st^{(1)}\rmv,\ldots,\st^{(n)}\}$ with $n \leq J$, the multiobject pdf $f^{\text{MBM}}(\RFSst)$ is given\vspace{-1mm} by
\begin{equation*}
\label{eq:fund3_2}
f^{\text{MBM}}(\RFSst) = \sum^{I}_{i\ist=\ist 1} w_i \ist f_i^{\text{MB}}(\RFSst)
\vspace{-1mm}
\end{equation*}
where $I$ is the number of MB pdfs $f_i^{\text{MB}}(\RFSst)$ and $\sum^{I}_{i\ist=\ist 1} \rmv w_i = 1$.

The cardinality of the \textit{Poisson RFS} $\RFSstR$ is Poisson distributed with mean $\mu$, i.e., $\rho(n) = \Set{P}(n;\mu) = \mathrm{e}^{-\mu}\mu^{n}/n!\ist$, $n \!\in\! \mathbb{N}_0$. For each cardinality value ${\sf n} \rmv=\rmv |\RFSstR|$, the individual elements $\stR$ are independent and identically distributed (iid) according to some spatial pdf $f(\st)$; thus $f_n(\st^{(1)}\rmv,\ldots,\st^{(n)}) = \prod_{i=1}^n f(\st^{(i)})$. 
According to \eqref{eq:fund0}, the multiobject pdf can be found as \cite{Mah07}
\vspace{0.5mm}
\begin{equation}
\label{eq:PoisPDF}
f^{\text{P}}(\RFSst) \ist=\ist \mathrm{e}^{-\int\PHD(\st')\ist\text{d}\st'}\prod_{\st \ist\in \RFSst} \rmv \PHD(\st)
\vspace{-.5mm}
\end{equation}
where $\PHD(\st) = \mu f(\st)$ is designated as \textit{PHD} or \textit{intensity function}.

A \textit{Poisson/MB (PMB) RFS} $\RFSstR$ is the union of a Poisson RFS and an MB RFS. 
The pdf of a PMB RFS can be obtained by applying the set convolution to the Poisson pdf and the corresponding MB pdf. 
This leads to the multiobject pdf $f^{\text{PMB}}(\RFSst)$ \vspace{0.5mm} given by
\begin{align}
f^{\text{PMB}}(\RFSst) \hspace{1mm}= \sum_{\RFSst^{(1)} \uplus\ist \RFSst^{(2)} =\ist\RFSst}\hspace{-3mm} f^{\text{P}}\big(\RFSst^{(1)}\big)\ist f^{\text{MB}}\big(\RFSst^{(2)}\big). \label{eq:fund4}\\[-5mm]
\nn
\end{align}

Finally, a \textit{Poisson/MBM (PMBM) RFS} $\RFSstR$ is the union of a Poisson RFS and an MBM RFS.
Similarly to the PMB RFS, the pdf of a PMBM RFS can be obtained by using the set convolution, which yields
\vspace{0.5mm}
\begin{align}
f^{\text{PMBM}}(\RFSst) \hspace{1mm}= \sum_{\RFSst^{(1)} \uplus\ist \RFSst^{(2)} =\ist\RFSst}\hspace{-3mm} f^{\text{P}}\big(\RFSst^{(1)}\big)\ist f^{\text{MBM}}\big(\RFSst^{(2)}\big). \label{eq:fund5}\\[-6mm]
\nn
\end{align}
Note that the sums in \eqref{eq:fund3}, \eqref{eq:fund4}, and \eqref{eq:fund5} ensure permutation invariance of the corresponding multiobject pdfs with respect to its arguments, but are never explicitly evaluated in the implementation of our proposed tracking method\vspace{-3.5mm}.

\vspace{3mm}
\section{Object States and State Transition Model}
\label{sec:sys}
\vspace{0mm} 

We model the multiobject state at time $k$ by an RFS $\RFSstR_{k} \rmv= \{\stR_{k}^{(1)}\rmv,\ldots,\stR_{k}^{(n_k)}\}$. 
The elements, i.e., the single-object state vectors $\stR_{k}^{(i)}$\rmv, $i \rmv\in\rmv \Set{I}_k\rmv \triangleq \rmv\{1,\ldots,n_k\}$, represent two-dimensional (2D) position and 2D velocity $\RV{p}_{k}^{(i)} = [\rv{p}_{k,1}^{(i)} \,\, \rv{p}_{k,2}^{(i)} \,\, \dot{\rv{p}}_{k,1}^{(i)} \,\, \dot{\rv{p}}_{k,2}^{(i)}]^{\T}$ and the object's intensity $\rv{\gamma}_k^{(i)}\rmv$, i.e., $\stR_k^{(i)} \!= [\RV{p}_{k}^{(i)\ist\T} \,\, \rv{\gamma}_k^{(i)}]^{\T}\rmv$.

The multiobject state transition model describes the statistics of the state $\RFSstR_{k}$ at time $k$ conditioned on the state $\RFSst_{k-1}$ at time $k-1$ and will be presented in the following. 
In fact, we model the state transition using the well-established model described in, e.g., \cite{Mah07,Wil:J15,Mey18Proc}.
Here, at time $k \rmv-\rmv 1$, an object with state $\st_{k-1} \!\in\rmv \RFSst_{k-1}$ survives with probability $\su(\st_{k-1})$ or dies with probability $1 \!-\rmv \su(\st_{k-1})$. If the object survives, its state $\stR_k$ at time $k$ is distributed according to the single-object state transition pdf $f(\st_k|\st_{k-1})$. 
We assume that the states of different objects evolve independently, i.e., given $\st_{k-1}$, $\stR_k$ is conditionally independent of all the other single-object states $\stR_k'$. 
Given these assumptions, the states of the survived objects $\RFSstR_k^{\text{S}}$, conditioned on $\RFSst_{k-1}$, is an MB RFS given by $\RFSstR_k^{\text{S}} \rmv= \bigcup_{\st_{k-1} \in\ist \RFSst_{k-1}} \!\RS{S}_k(\st_{k-1})$. 
Here, the Bernoulli component $\RS{S}_k(\st_{k-1})$ is parametrized by (cf. Eq.~\eqref{eq:fund2}) the existence probability $\su(\st_{k-1})$ and the spatial pdf $f(\st_k|\st_{k-1})$. 

There are also newborn objects, which are modeled by the Poisson RFS $\RFSstR_{k}^{\text{B}}\rmv$ with cardinality mean $\mu_{\text{B}}$, spatial pdf $f_{\text{B}}(\st_k)$, and hence PHD $\lambda_{\text{B}}(\st_k) \rmv= \mu_{\text{B}} \ist f_{\text{B}}(\st_k)$. Conditioned on $\RFSst_{k-1}$, survived objects $\RFSstR_k^{\text{S}}$ and newborn objects $\RFSstR_k^{\text{B}}$ are independent. Thus, for $\RFSst_{k-1}$ fixed, the overall multiobject state at time $k$,  is given \vspace{-1.5mm}by 
\begin{equation}
\label{eq:STpdf}
\RFSstR_k = \RFSstR_k^{\text{S}} \cup \RFSstR_k^{\text{B}} 
  = \Bigg( \bigcup_{\st_{k-1} \in\ist \RFSst_{k-1}} \!\!\!\RS{S}_k\big(\st_{k-1}\big) \rmv \Bigg) \cup \RFSstR_k^{\text{B}}\ist.
\vspace{0.5mm}
\end{equation}
The state transition pdf $f(\RFSst_k|\RFSst_{k-1})$ corresponding to \eqref{eq:STpdf} can be found in, e.g., \cite{Mah07,Wil:J15,Mey18Proc}.


\section{Measurement Model}
\label{sec:measmod}

The measurement model describes the statistics of the measurement $\mevR_k$ conditioned on the state $\RFSst_k \rmv= \{\st_k^{(1)}\rmv,\ldots,\st_k^{(n_k)}\}$.
Here, $\mevR_k$ represents data cells and is modeled by a random vector according to $\mevR_k \triangleq [\meR_k^{(1)}\rmv\ldots\ist\meR_k^{(M)}]^{\T}$\rmv, where each entry $\meR_k^{(m)}$\rmv, $m \in \mathcal{M} \triangleq \{1,\ldots,M\}$ in turn represents a scalar intensity value of the corresponding data cell~ $m$.
The statistics of $\mevR_k$, conditioned on $\RFSst_k$, will be described in the following\vspace{-3.5mm}.


\subsection{Association-Based Data Cell Model}
\label{sec:MeaGen0}
\vspace{-.5mm}

In what follows, we present the association-based data cell model used by our proposed multiobject tracking method. 
The key assumptions of our model are as follows\vspace{.5mm}.
\begin{enumerate}
\item An object $\st_k$ can only contribute to data cell $m$ if it is located in it\vspace{.5mm}.
\item If there is only a single object $\st_k$ located in data cell~$m$, it must contribute to it\vspace{.5mm}.
\item If there are multiple objects $\st_k^{(i)}$\rmv, $i \rmv\in\rmv \Set{I}_k$, located in data cell~$m$, only one of them contributes to it\vspace{.5mm}.
\end{enumerate}
Assumption 3) is in contrast to superpositional data cell models \cite{Lia23TBD,Dav24TBD}, but, as demonstrated in Section~\ref{sec:num}, can lead to a very good tracking performance even if the true data model in force is superpositional. 
Assumptions 2) and 3) lead to the derivation of object contribution probabilities in Section~\ref{sec:MeaGen1}.


To describe unknown object-to-cell associations, we introduce the random association vector $\assvR_k \rmv\triangleq [\assR_k^{(1)}\rmv\ldots\ist\assR_k^{(n_k)}]^{\T}$ with entries $\assR_k^{(i)} \rmv\rmv\in\rmv\rmv \Set{M}$. 
Here, $\assR_k^{(i)} \rmv=\rmv m \rmv\in\rmv \Set{M}$ indicates that object $\st_k^{(i)}$ is located in data cell $m$. 
We collect all possible association vectors $\assv_k$ in the set $\mathcal{A}_k \rmv\triangleq\rmv \Set{M}^{n_k}$. 
For\vspace{.3mm} future reference, we also introduce the sets 
$\mathcal{M}_{\V{a}_k} \rmv\triangleq\rmv \big\{a^{(i)}_k\big\}_{i\ist\in\ist\Set{I}_k}$
and $\Set{I}_{\assv_k}^{(m)} \triangleq \big\{ i\in\Set{I}_k \ist | \ist a^{(i)}_k = m \big\}$ as well as $n_k^{(m)} \rmv\triangleq\rmv |\Set{I}_{\assv_k}^{(m)}|$.



To further describe which object contributes to a data cell occupied by multiple objects, we introduce the random binary vector $\RV{\theta}_k$. 
Here, $\RV{\theta}_k$ is defined as $\RV{\theta}_k \rmv\triangleq [\rv{\theta}_k^{(1)}\rmv\ldots\ist\rv{\theta}_k^{(n_k)}]^{\T}$, where each entry $\rv{\theta}_k^{(i)} \rmv\in\rmv\{0,1\}$ indicates whether object $\st_k^{(i)}$ contributes to a data cell \big($\rv{\theta}_k^{(i)} \rmv=\rmv 1$\big) or not \big($\rv{\theta}_k^{(i)} \rmv=\rmv 0$\big). 
For a given $\assv_k\rmv\in\rmv\mathcal{A}_k$, we introduce the set $\Theta_{\assv_k}$, which consists of all vectors $\V{\theta}_k$ that have exactly one entry $\theta_k^{(i)}\rmv = 1$ for all $i \in \Set{I}_{\assv_k}^{(m)}$\rmv.

For example, lets consider the case where $M = 4$, $n_k = 3$, and $\assv_k = [2 \ist\ist\ist 3 \ist\ist\ist 1]^\T\rmv\rmv$. Here, $\assv_k$ indicates that each of the three objects with indexes $i\rmv\in\rmv\{1,2,3\}$ is in a different data cell, i.e., $a^{(i)}_k \rmv\rmv\neq\rmv\rmv a^{(j)}_k\rmv\rmv$, $\forall i,j \in \Set{I}_k $ with $i \neq j$. 
Accordingly, the set $\Theta_{\assv_k}$ consists of only one binary vector, which is given by $\V{\theta}_k = [1 \ist\ist\ist 1 \ist\ist\ist 1]^\T$. 
On the other hand, let now 
$\assv_k = [2 \ist\ist\ist 3 \ist\ist\ist 2]^\T\rmv\rmv$. Now $\assv_k$ indicates that the objects with indexes $i\rmv=\rmv1$ and $i\rmv=\rmv3$ are in the same data cell $m = 2$. 
In this case, the set $\Theta_{\assv_k}$ consists of two binary vectors, which are given by $\V{\theta}_k = [1 \ist\ist\ist 1 \ist\ist\ist 0]^\T$ and $\V{\theta}_k  = [0 \ist\ist\ist 1 \ist\ist\ist 1]^\T\rmv\rmv$. Here, $\V{\theta}_k \rmv=\rmv [1 \ist\ist\ist 1 \ist\ist\ist 0]^\T$ indicates that object $i\rmv=\rmv1$ contributes to data cell $m=2$, while $\V{\theta}_k  = [0 \ist\ist\ist 1 \ist\ist\ist 1]^\T\rmv\rmv$ indicates that object $i\rmv=\rmv3$ contributes to data cell $m\rmv=\rmv2$.

In summary, the object-to-cell association vector $\assvR_k$ in combination with the object contribution vector $\RV{\theta}_k$ allows us to define all possible object-to-cell contributions. 
In particular, $\assR_k^{(i)} \rmv\rmv\in\rmv\rmv \Set{M}$ defines in which data cell object $\st_k^{(i)}$ is located and $\rv{\theta}_k^{(i)} \rmv\in\rmv \{0,1\}$ defines whether it contributes to that cell or not.
If object $\st_k^{(i)}$ contributes to data cell $m$, $\meR_k^{(m)}$ is distributed according to $f_{1}\big(\me^{(m)}_k\big|\st_k^{(i)}\big)$. 
If no object is located in data cell $m$, $\meR_k^{(m)}$ is purely determined by noise and is distributed according to $f_{0}\big(\me^{(m)}_k \big)$\vspace{-1mm}.


\subsection{The Joint Likelihood Function}
\vspace{-.5mm}
\label{sec:JLF}

Based on $\assvR_k$ and $\RV{\theta}_k$, the joint likelihood function $f(\mev_k|\RFSst_k)$ can now be computed by using the total probability theorem according to
\vspace{0mm}
\begin{align}
f(\mev_k|\RFSst_k) \ist&=\rmv \sum_{\assv_k \in\ist \mathcal{A}_{k}}\ist \sum_{\V{\theta}_k \in\ist \Theta_{\assv_k}} f(\mev_k|\assv_k,\rmv\V{\theta}_k,\rmv\RFSst_k)\iist p(\assv_k,\rmv\V{\theta}_k|\RFSst_k)\ist. 
\nn \\[-4mm]
\label{eq:LikeBinom_1} \\[-5.5mm]
\nn
\end{align}	
In what follows, we present expressions for the conditional distributions $f(\mev_k|\assv_k,\V{\theta}_k,\RFSst_k)$ and $p(\assv_k,\V{\theta}_k|\RFSst_k)$ based on the data cell model proposed in the previous subsection.

By further assuming that, conditioned on $\assv_k$, $\V{\theta}_k$, and $\RFSst_k$, all cell measurements $\me_k^{(m)}$\rmv, $m\rmv\in\rmv\Set{M}$ are statistically independent among each other, the conditional pdf $f(\mev_k|\assv_k,\V{\theta}_k,\RFSst_k)$ can be found\vspace{0mm} as
\vspace{0mm}
\begin{align}
&f(\mev_k|\assv_k,\rmv\V{\theta}_k,\rmv\RFSst_k) \nonumber\\[1mm]
&\hspace{8mm}= \Big(\prod_{i \ist\in\ist \Set{I}_{\V{\theta}_k}}\rrmv f_1\big(\me_k^{(\ass_k^{(i)})}\big|\st_k^{(i)}\big) \Big) \rmv\rmv\rmv \prod_{m \ist\in\ist \mathcal{M} \backslash \mathcal{M}_{\V{a}_k}}\rmv\rmv\rmv\rmv f_0\big(\me_k^{(m)}\big). \label{eq:IntensityModel} \\[-5.5mm]
\nn
\end{align}	
Here, the set $\Set{I}_{\V{\theta}_k}$ consists of all object indexes $i\rmv\in\rmv\Set{I}_k$ with $\theta_k^{(i)} \rmv= 1$. 

Next, we can factorize the prior object-to-cell association pmf $p(\assv_k,\rmv\V{\theta}_k|\RFSst_k)$ according to
\vspace{1mm}
\begin{equation}
\label{eq:Lim4}
p(\assv_k,\rmv\V{\theta}_k|\RFSst_k) =\ist p(\V{\theta}_k|\assv_k,\rmv\RFSst_k) \ist p(\assv_k|\RFSst_k) \ist. 
\end{equation}
Here, the association pmf $p(\assv_k|\RFSst_k)$ is defined as 
\vspace{0mm}
\begin{equation}
\label{eq:prior1}
p(\assv_k|\RFSst_k) = \prod_{i\ist\in\ist\Set{I}_{k}} \delta^{(\ass_k^{(i)})}(\st_k^{(i)}) 
\vspace{-0.5mm}
\end{equation}
with the indicator function $\delta^{(m)}(\st_k)$  given by
\vspace{0mm}
\begin{equation}
\label{eq:delta1}
\delta^{(m)}(\st_k) = 
\begin{cases}
1, & \st_k\iist \text{is in data cell} \iist m \\[0mm]
0, & \st_k\iist \text{is not in data cell} \iist m\ist.
\end{cases}
\vspace{0.5mm}
\end{equation}
Note that the definition of $p(\assv_k|\RFSst_k)$ in \eqref{eq:prior1} and \eqref{eq:delta1} enforces Assumption 1 in Section \ref{sec:MeaGen0}.
Finally, we introduce the \textit{object contribution probabilities}
\vspace{1mm}
\begin{align}
p(\V{\theta}_k|\assv_k,\rmv\RFSst_k) \ist&=\rmv \prod_{m \ist\in\ist \Set{M}_{\ass_k}}\rmv\rmv p^{(m)}\big(\V{\theta}^{(m)}_{k}|\RFSst_{k}\big)\ist. \label{eq:prior2} \\[-6mm]
\nn
\end{align}
Here, the vector $\V{\theta}^{(m)}_{k} \rmv= \big[\theta_k^{(m,1)}\rmv\ldots\ist\theta_k^{(m,n_k^{(m)})}\big]^{\T}$ comprises all entries $\theta_k^{(i)}$ of $\V{\theta}_k$ with $i \rmv\in\rmv  \Set{I}_{\assv_k}^{(m)}$ in an arbitrary but fixed order.
Inserting \eqref{eq:prior1} and \eqref{eq:prior2} into \eqref{eq:Lim4}\vspace{1mm} yields
\begin{align}
&\hspace{-3mm}p(\assv_k,\rmv\V{\theta}_k|\RFSst_k) \nn\\[1.5mm]
&=\ist \Big(\prod_{i\ist\in\ist\Set{I}_{k}} \delta^{(\ass_k^{(i)})}(\st_k^{(i)}) \Big) \prod_{m \ist\in\ist \Set{M}_{\ass_k}}\rmv p^{(m)}\big(\V{\theta}^{(m)}_{k}|\RFSst_{k}\big). \label{eq:prior3} \end{align}

\subsection{Object Contribution Probabilities}
\label{sec:MeaGen1}
\vspace{0mm}

The object contribution probabilities $p^{(m)}\big(\V{\theta}^{(m)}_{k}|\RFSst_{k}\big)$ are obtained based on concepts from classical detection theory \cite{Kay:B93}.
More precisely, we first propose object contribution probabilities for the case of \emph{thresholded data cells} described by $p_{\eta}^{(m)}\big(\V{\theta}^{(m)}_{k}|\RFSst_{k}\big)$, which are based on the detection threshold~$\eta$. As we will derive later in this section, an expression for $p^{(m)}\big(\V{\theta}^{(m)}_{k}|\RFSst_{k}\big)$ can then be obtained by letting $\eta$ approach zero. For future reference, we introduce the set of indexes $\Set{I}_k^{(m)} = \big\{1, \dots, n_k^{(m)}\big\}$.

Let us first consider the case where a single object $\st_k^{(i)}$ is located in data cell $m$. 
Following classical detection theory, the probability that object $\st_k^{(i)}$ is detected, i.e., $\meR_k^{(m)} \rmv\geq \eta$, is given\vspace{0mm} by
\begin{align}
p_{\eta}^{(i,m)} &\triangleq\ist \int_{\eta}^{\infty} \rmv\rmv f_1\big(\me_k^{(m)}\big|\st_k^{(i)}\big)\ist \text{d}\me_k^{(m)} \label{eq:Pd1}
\end{align} 
where $f_1\big(\me_k^{(m)}\big|\st_k^{(i)}\big)$ is the conditional pdf of $\meR_k^{(m)}$ given $\st_k^{(i)}$ (cf. Section \ref{sec:MeaGen0}). Thus, for the case where a single object is in the cell $m$, $p_{\eta}^{(i,m)}$ is the probability that the object is detected and $1 - p_{\eta}^{(m,i)}$ is the probability that the object is not detected. 

We now define our model for the object contribution probabilities based on the probabilities of detection discussed above and Assumption 2) and Assumption 3) from Section~\ref{sec:MeaGen0}. In particular, for $i \in \Set{I}_k^{(m)}$\rmv, we define 
\begin{align}
&\hspace{-2mm}p^{(m)}_{\eta}\big(\V{\theta}^{(m)}_{k} \rmv\rmv=\rmv \V{\theta}_k^{(m,i)}|\Set{X}_{k}\big) \nn \\[1mm]
&\hspace{8mm} = \frac{p_{\eta}^{(i,m)} \prod_{i'\ist\in\ist\Set{I}_{k}^{(m)}\setminus \{i\}} \big( 1 - p_{\eta}^{(i',m)}  \big)}{C^{(m)}_{\eta}} \label{eq:priorEta2}  \\[-5mm]
\nn
\end{align}
where the vector $\V{\theta}_k^{(m,i)}$ is defined as the $i$'s entry of $\V{\theta}_k^{(m)}$ being one, i.e., $\theta_k^{(m,i)}\rmv = 1$, and all the other entries being zero, i.e., $\theta_k^{(m,i')}\rmv = 0$ for $i' \in \Set{I}_k^{(m)} \rmv\setminus\rmv \{i\}$.
The normalization factor in \eqref{eq:priorEta2} can be obtained by summing over all possible numerators in \eqref{eq:priorEta2}, i.e.,
\vspace{1mm}
\begin{align}
C^{(m)}_{\eta} \ist=\ist \hspace{-2mm} \sum_{i\ist\in\ist\Set{I}_{k}^{(m)}} \hspace{0mm}  p_{\eta}^{(i,m)} \hspace{-1mm} \prod_{i'\ist\in\ist\Set{I}_{k}^{(m)}\setminus \{i\}}  \hspace{-1mm}  \big( 1 - p_{\eta}^{(i'\rmv,m)} \big).  \label{eq:priorConstant} \\[-5.5mm]
\nn
\end{align}
Note that for the case with a single object, i.e.,  $\Set{I} _{k}^{(m)} \rmv=\rmv \{ i \}$,  $p^{(m)}_{\eta}\big(\V{\theta}^{(m)}_{k} \rmv\rmv=\rmv \V{\theta}_k^{(m,i)}|\Set{X}_{k}\big)  = 1$.

After having defined $p_{\eta}^{(m)}\big( \V{\theta}^{(m)}_{k} | \ist \Set{X}_{k} \big)$, we finally obtain the object contribution pmf $p^{(m)}\big( \V{\theta}^{(m)}_{k} | \ist \Set{X}_{k} \big)$ by letting the threshold~$\eta$ approach zero, i.e.,
\vspace{0.5mm}
\begin{equation}
p^{(m)}\big( \V{\theta}^{(m)}_{k} | \ist \Set{X}_{k} \big) = \lim_{\eta\ist\rightarrow\ist 0} \ist\ist p_{\eta}^{(m)}\big( \V{\theta}^{(m)}_{k} | \ist \Set{X}_{k} \big) .  
\label{eq:probDet}
\end{equation}
Now that we have formulated our general modeling approach for $p^{(m)}\big( \V{\theta}^{(m)}_{k} | \ist \Set{X}_{k} \big)$, let us next elaborate it in terms of the concrete example of Swerling 1 objects, which are often assumed in radar applications \cite{Ris19,Lep16}\vspace{-2mm}.

\subsection{Example: Swerling 1 Objects}
\label{sec:MeaGen2}
\vspace{-.5mm}

%
%

In many radar applications, a standard model for $\meR_k^{(m)}$ is given by \cite{Kim21}
\vspace{0.5mm}
\begin{equation}
\label{eq:Meas1}
\meR_k^{(m)} = \left|\sum_{i \ist\in\ist \Set{I}_k} \rv{\rho}(\st_k^{(i)})\ist \sqrt{d^{(m)}(\st_k^{(i)})} + \rv{w}_k^{(m)}\right|\ist.
\vspace{-1mm}
\end{equation} 
Here, object $\st_k^{(i)}$\rmv, $i \rmv\in\rmv \Set{I}_k$ contributes to cell intensity $\meR_k^{(m)}$ via the deterministic factor $\sqrt{d^{(m)}(\st_k^{(i)})}$ and the random factor $\rv{\rho}(\st_k^{(i)})$, where $d^{(m)}(\st_k^{(i)})$ is usually referred to as point spread function (PSF). 
In addition, $\rv{w}_k^{(m)}$ is an object-independent additive complex noise component, which is assumed to be iid across time $k$ and cell index $m$ and is described by the pdf $f(w_k^{(m)})$.
In this general setting, cell intensity $\meR_k^{(m)}$ can have contributions from all objects $\st_k^{(i)}$\rmv, $i \rmv\in\rmv \Set{I}_k$, and object $\st_k^{(i)}$ can contribute to all data cells $\meR_k^{(m)}$\rmv. 
However, our modeling approach assumes that there is at most one object contribution to a single data cell~$m$, which simplifies \eqref{eq:Meas1} to
\begin{equation}
\label{eq:Meas1_3}
\meR_k^{(m)} \ist=\ist \Big|\rv{\rho}(\st_k^{(i)})\ist \sqrt{d^{(m)}(\st_k^{(i)})} + \rv{w}_k^{(m)}\Big|\ist.
\vspace{0mm}
\end{equation}
Note that, for $n_k^{(m)} \rmv=\rmv 0$, expression \eqref{eq:Meas1_3} further simplifies to $\meR_k^{(m)} = \big|\rv{w}_k^{(m)}\big|$.

In the Swerling 1 case, the fluctuation of object returns $\rv{\rho}^{(m)}(\st_k^{(i)})$ and the background noise $\rv{w}_k^{(m)}$ are both modeled as independent and circularly symmetric zero-mean complex Gaussian with variance one and $\sigma^{2}_{0}$, respectively. 
It can then be shown \cite{Lep16} that under this assumption $f_0\big(\st_k^{(i)}\big)$ and $f_1\big(\me_k^{(m)}\big|\st_k^{(i)}\big)$ 
are Rayleigh distributed and given by 
\begin{align*}
f_0\big(\me_k^{(m)}\big) &= \Set{R}\big(\me^{(m)}_k; \sigma_{0}\big) \\[1mm]
f_1\big(\me_k^{(m)}\big|\st_k^{(i)}\big) &= \Set{R}\big(\me^{(m)}_k; \sigma_{m}(\st_k^{(i)}) \big) 
\end{align*}
respectively, with $\sigma_{m}(\st_k^{(i)}) \triangleq \sqrt{d(\st^{(i)}_k) + \sigma^{2}_{0}}$. 
Next, we insert $f_1\big(\me_k^{(m)}\big|\st_k^{(i)}\big)$ into \eqref{eq:Pd1}, which yields
\begin{equation}
\label{eq:Det3}
p_{\eta}^{(i,m)} =\ist \mathrm{e}^{\hspace{-2mm}-\frac{1}{2}\frac{\eta^2}{\sigma_{m}^2\left(\st_k^{(i)}\right)}} \ist.
\vspace{.5mm}
\end{equation}

In order to illustrate the application of \eqref{eq:probDet}, let us consider the case with $n^{(m)}_k \rmv\rmv\rmv=\rmv\rmv\rmv 2$ objects being in data cell $m$, i.e.,
$\st_k^{(i)}$ with $i \rmv\in \{1,2\}$. We now compute $p^{(m)}\rmv\big(\theta_k^{(m,1)}\rmv, \theta_k^{(m,2)}|\{\st_k^{(1)}\rmv\rmv, \st_k^{(2)}\} \big)$ for $\theta_k^{(m,1)}\rmv= 1$ and $\theta_k^{(m,2)} \rmv= 0$ according to \eqref{eq:probDet}. 
However, this first requires the specification of $p^{(m)}_{\eta}\big(1,\rmv 0\ist|\ist\{\st_k^{(1)}\rmv\rmv, \st_k^{(2)}\}\big)$ according to \eqref{eq:priorEta2}, which can be found as 
\vspace{0mm}
\begin{equation}
\label{priorEta2}
p^{(m)}_{\eta}\big(1\ist,\rmv 0|\{\st_k^{(1)}\rmv\rmv, \st_k^{(2)}\}  \big) =  \frac{p_{\eta}^{(1,m)}\ist p_{\eta}^{(2,m)}}{C^{(m)}_{\eta}}
\vspace{-1mm}
\end{equation}
with
\[
C^{(m)}_{\eta} =\ist p_{\eta}^{(1,m)} \big(1-p_{\eta}^{(2,m)}\big) + \big(1-p_{\eta}^{(1,m)}\big)  p_{\eta}^{(2,m)}\ist.
\]
By inserting \eqref{eq:Det3} into \eqref{priorEta2} and applying the limit $\eta \rmv\to\rmv 0$ according to \eqref{eq:probDet}, we see that we face a ``zero-by-zero'' division. 
However, this indeterminate limit can still be computed by applying the rule of de L'Hospital \cite{Kra04}. In particular, as we show in Appendix \ref{sec:App_Limit}, we get  
\[
p^{(m)}\big(1\ist,\rmv 0|\{\st_k^{(1)}\rmv\rmv, \st_k^{(2)}\} \big) = \frac{\sigma_{m}^{2}(\st_k^{(1)})}{\sigma_{m}^{2}(\st_k^{(1)})+\sigma_{m}^{2}(\st_k^{(2)})}\ist.
\vspace{.3mm}
\]
An analogous expression can be obtained for $\theta_k^{(m,1)}\rmv= 0$, $\theta_k^{(m,2)} \rmv= 1$.

The described procedure can be generalized to the case where there is an arbitrary number of objects in data cell~$m$. 
As further shown in Appendix~ \ref{sec:App_Limit}, for this general Swerling~1 case, we get
\vspace{-0.5mm}
\begin{equation}
p^{(m)}\rmv\big(\V{\theta}^{(m)}_{k} \rmv= \V{\theta}_k^{(m,i)}|\Set{X}_{k} \big) = \frac{\sigma^{2}_m(\st_k^{(i)})}{\sum_{i'=1}^{n^{(m)}_k}\sigma^{2}_m(\st_k^{(i')})}.\label{eq:priorEtaSwerling}
\vspace{1.5mm}
\end{equation}
Interestingly, in this concrete example where multiple Swerling~1 objects share the same cell, object association probabilities are nonzero and highly intuitive.

\section{Prediction Step}
\label{sec:pred}

To derive our proposed algorithm, we perform sequential Bayesian estimation, which requires the specification of a multiobject state model, a state transition model (cf. Section~\ref{sec:sys}), and a measurement model (cf. Section~\ref{sec:measmod}). 
The multiobject state $\RFSstR_k$ at time $k$, given all collected measurements $\mev_{1:k} \triangleq (\mev_1,\ldots,\mev_k)$ up to time $k$, is described by the posterior pdf $f(\RFSst_k|\mev_{1:k})$.
Here, we model the multiobject state $\RFSstR_{k-1}$ at time $k\rmv-\rmv 1$ by a PMB RFS. 
Thus, the posterior pdf $f(\RFSst_{k-1}|\mev_{1:k-1})$ at time $k\rmv-\rmv 1$ is given as (cf. Eq.~\eqref{eq:fund4})
\vspace{0mm}
\begin{align}
f(\RFSst_{k-1}|\mev_{1:k-1}) \ist&=\! \sum_{\RFSst_{k-1}^{(1)} \uplus\ist \RFSst_{k-1}^{(2)} =\ist\RFSst_{k-1}}\hspace{-3mm} f^{\text{P}}\big(\RFSst_{k-1}^{(1)}\big)\ist f^{\text{MB}}\big(\RFSst_{k-1}^{(2)}\big) \nn \\[-3mm]
\label{eq:pred1} \\[-5.5mm]
\nn
\end{align}
where $f^{\text{P}}\big(\RFSst_{k-1}\big)$ is a Poisson pdf (cf. Eq. \eqref{eq:PoisPDF}) fully characterized by the PHD $\lambda(\st_{k-1})$, and $f^{\text{MB}}\big(\RFSst_{k-1}\big)$ is an MB pdf (cf. Eq. \eqref{eq:fund3}) consisting of $J_{k-1}$ Bernoulli components fully characterized by the existence probabilities $\ex_{k-1}^{(j)}$ and spatial pdfs $f^{(j)}(\st_{k-1})$, $j \rmv\in\rmv \Set{J}_{k-1} \triangleq\{1,\ldots,J_{k-1}\}$. 
Sequential Bayesian estimation now constitutes recursively computing $f(\RFSst_k|\mev_{1:k})$ by means of a prediction step and an update step.
While the update step will be presented in the next section, the prediction step is presented in the following.

The prediction step converts the posterior pdf $f(\RFSst_{k-1}|\mev_{1:k-1})$ at time $k\rmv-\rmv 1$ to the predicted posterior pdf $f(\RFSst_{k}|\mev_{1:k-1})$ at time $k$ using the state transition pdf $f(\RFSst_k|\RFSst_{k-1})$, which is based on the state transition model described in Section \ref{sec:sys}.
In fact, the prediction step for the PMB prior in \eqref{eq:pred1} and the state transition model of Section~ \ref{sec:sys} coincides with the prediction step of the conventional PMB filter for point measurements \cite{Wil:J15,Kro16,Gar18PMBM}.
Hence, we will only briefly review it in the following.

As was shown in \cite{Wil:J15}, the prediction step preserves the PMB form in \eqref{eq:pred1}, which means that the predicted posterior pdf $f(\RFSst_k|\mev_{1:k-1})$ is again of PMB form. In fact, we have
\vspace{0.5mm}
\begin{align}
\hspace{-1.5mm} f(\RFSst_{k}|\mev_{1:k-1}) &=\hspace{-2mm} \sum_{\RFSst_k^{(1)} \uplus\ist \RFSst_k^{(2)} =\ist\RFSst_{k}}\hspace{-3mm} f_{k|k-1}^{\text{P}}\big(\RFSst_k^{(1)}\big)\ist f_{k|k-1}^{\text{MB}}\big(\RFSst_k^{(2)}\big) \ist. \label{eq:pred1_2}\\[-5mm]
\nn \\[-4.5mm]
\nn
\end{align}
The Poisson pdf $f_{k|k-1}^{\text{P}}\big(\RFSst_k\big)$ is characterized by the PHD $\lambda_{k|k-1}(\st_k)$, which is given by 
\begin{align}
&\hspace{-1mm}\lambda_{k|k-1}(\st_k)  \nn \\
& \hspace{1mm}=\lambda_{\text{B}}(\st_k) +\rmv \int\! p_{\text{S}}(\st_{k-1}) f(\st_k|\st_{k-1}) \ist \lambda(\st_{k-1}) \ist \mathrm{d}\st_{k-1} \ist. \label{eq:pred2} 
\end{align}
Here, $\lambda_{\text{B}}(\st_k)$, $p_{\text{S}}(\st_{k-1})$, and $f(\st_k|\st_{k-1})$ are the PHD representing newborn objects, the survival probability, and the single-object state transition pdf, respectively, introduced in Section \ref{sec:sys}, and $\lambda(\st_{k-1})$ is the PHD characterizing the Poisson pdf $f^{\text{P}}\big(\RFSst_{k-1}\big)$ in \eqref{eq:pred1}.
Furthermore, the MB pdf $f_{k|k-1}^{\text{MB}}\big(\RFSst_k\big)$ consists of $J_{k-1}$ Bernoulli pdfs with existence probabilities and spatial pdfs given by 
\begin{align}
\ex_{k|k-1}^{(j)} &=\ist \ex_{k-1}^{(j)} \!\int \! p_{\text{S}}(\st_{k-1}) \ist f^{(j)}(\st_{k-1}) \ist \mathrm{d}\st_{k-1} \label{eq:pred3}\\[1.5mm]
f_{k|k-1}^{(j)}(\st_k) &\ist=\ist \frac{\int p_{\text{S}}(\st_{k-1})\ist f(\st_k|\st_{k-1}) \ist f^{(j)}(\st_{k-1}) \ist 
  \mathrm{d}\st_{k-1}}{\int \rmv p_{\text{S}}(\st'_{k-1}) \ist f^{(j)}(\st'_{k-1}) \ist \mathrm{d}\st'_{k-1}} \label{eq:pred4}
\\[-4.5mm]
\nn
\end{align}
respectively, with $j \rmv\in\rmv \Set{J}_{k-1}$. 
Here, $\ex_{k-1}^{(j)}$ and $f^{(j)}(\st_{k-1})$ are the parameters characterizing the MB pdf $f^{\text{MB}}\big(\RFSst_{k-1}\big)$ in \eqref{eq:pred1}. 
Note that the number of Bernoulli components $J_{k-1}$ is not changed by the prediction step. 
Newborn objects are modeled by the PHD $\lambda_{\text{B}}(\st_k)$ in \eqref{eq:pred2}\vspace{-4mm}.

\vspace{2mm}
\section{Update Step - Stage One}
\label{sec:upd_step}

In the following two sections, we perform the update step based on our association-based data cell model proposed in Section \ref{sec:measmod}.
This involves first an exact update step, followed by a series of modifications and approximations leading to the overall update step of our proposed PMB tracking method\vspace{0mm}. 

\subsection{Exact Update Step}
\label{sec:upd_ex}
\vspace{0.5mm}

In the update step, the predicted posterior pdf $f(\RFSst_k|\mev_{1:k-1})$ is converted into the new posterior pdf $f(\RFSst_{k}|\mev_{1:k})$ using the likelihood function $f(\mev_k|\RFSst_k)$ and the measurements $\mev_k$ obtained at time $k$. 
In fact, using Bayes' theorem, we have 
\vspace{0.5mm}
\begin{equation}
\label{eq:UpdEx1}
f(\RFSst_k|\mev_{1:k}) \propto f(\mev_k|\RFSst_k)\ist f(\RFSst_k|\mev_{1:k-1}) \ist.
\vspace{0.5mm}
\end{equation}
Inserting the expression of the likelihood function $f(\mev_k|\RFSst_k)$ of our association-based data cell model given by \eqref{eq:LikeBinom_1} into \eqref{eq:UpdEx1} and grouping terms yields 
\vspace{1mm}
\begin{align}
f(\RFSst_k|\mev_{1:k}) &\propto\rmv \sum_{\assv_k \in\ist \mathcal{A}_{k}} \sum_{\V{\theta}_k \in\ist \Theta_{\assv_k}} f(\mev_k|\assv_k,\rmv\V{\theta}_k,\rmv\RFSst_k) \nn \\[-1.5mm]
& \hspace{27mm}\times f(\assv_k,\rmv\V{\theta}_k,\rmv\RFSst_k|\mev_{1:k-1}) \ist. \label{eq:UpdEx2} \\[-4mm]
\nn
\end{align}
Here, the pdf $f(\mev_k|\assv_k,\rmv\V{\theta}_k,\rmv\RFSst_k)$ is given by \eqref{eq:IntensityModel} 
and the distribution $f(\assv_k,\rmv\V{\theta}_k,\rmv\RFSst_k|\mev_{1:k-1})$ by
\vspace{0.5mm}
\begin{equation}
\label{eq:UpdEx3} 
f(\assv_k,\rmv\V{\theta}_k,\rmv\RFSst_k|\mev_{1:k-1}) =\ist p(\assv_k,\rmv\V{\theta}_k|\RFSst_k)\ist f(\RFSst_k|\mev_{1:k-1})
\vspace{1mm}
\end{equation}
with $p(\assv_k,\rmv\V{\theta}_k|\RFSst_k)$ given by \eqref{eq:Lim4} and $f(\RFSst_k|\mev_{1:k-1})$ by \eqref{eq:pred1_2}.
Note that the posterior pdf $f(\RFSst_k|\mev_{1:k})$ in \eqref{eq:UpdEx2} is no longer a PMB pdf, but a pdf of a more complex form.
In the following, we apply a series of judiciously chosen approximations in order to convert the posterior pdf back into a PMB pdf.

As a first preparatory step, we rewrite the posterior pdf $f(\RFSst_k|\mev_{1:k})$ in \eqref{eq:UpdEx2} in terms of the \emph{augmented association vector} $\bar{\assvR}_k$ and the \emph{augmented object contribution vector} $\bar{\RV{\theta}}_k$.
In fact, we define $\bar{\assvR}_k \rmv\rmv\triangleq\rmv [\bar{\assR}_k^{(1)}\rmv\ldots\ist\bar{\assR}_k^{(J_k)}]^{\T}$ and $\bar{\RV{\theta}}_k \rmv\triangleq [\rv{\theta}_k^{(1)}\rmv\ldots\ist\rv{\theta}_k^{(J_k)}]^{\T}$ with $J_k = J_{k-1} \rmv+\rmv M$.
Here, $\bar{\assR}_k^{(j)} \rmv\in \rmv\{0\} \cup \Set{M}$ for $j \rmv\in\rmv \Set{J}_{k-1}$ 
and $\bar{\assR}_k^{(j)} \in \{0,m\}$ for $j \rmv=\rmv J_{k-1}\rmv+\rmv m$ with $m\in\Set{M}$. 
Furthermore, $\rv{\theta}_k^{(j)} \rmv\in\rmv \{0,1\}$ for all $j \rmv\in\rmv \Set{J}_k$. 
Note that $\Set{J}_k = \Set{J}_{k-1} \rmv\cup \Set{J}_k^{\text{N}}$ with $\Set{J}_k^{\text{N}} \triangleq \{J_{k-1}\rmv+\rmv 1,\ldots,J_{k-1}\rmv+\rmv M\}$.
The interpretation of the specific entries $[\bar{\assR}_k^{(j)}\ist \rv{\theta}_k^{(j)}]^{\T}$ is as follows: for $j \rmv\in\rmv \Set{J}_{k-1}$, $[\bar{\assR}_k^{(j)}\ist \rv{\theta}_k^{(j)}]^{\T} \rmv=\rmv [0\iist 0]^{\T}$
indicates that the object modeled by Bernoulli component~$j$ does not exist;
$[\bar{\assR}_k^{(j)}\ist \rv{\theta}_k^{(j)}]^{\T} \rmv=\rmv [m\iist 0]^{\T}$ with $m\in\Set{M}$ that the object exists, is located in but does not contribute to data cell~$m$;
and $[\bar{\assR}_k^{(j)}\ist \rv{\theta}_k^{(j)}]^{\T} \rmv=\rmv [m\iist 1]^{\T}$ that the object exists, is located in and also contributes to data cell~$m$.
For $j \rmv\in\rmv \Set{J}_k^{\text{N}}$, $[\bar{\assR}_k^{(j)}\ist \rv{\theta}_k^{(j)}]^{\T} \rmv=\rmv [m\iist 0]^{\T}$
indicates that an object modeled by the Poisson RFS is located in but does not contribute to data cell~$m$;
 and $[\bar{\assR}_k^{(j)}\ist \rv{\theta}_k^{(j)}]^{\T} \rmv=\rmv [m\iist 1]^{\T}$ that an object modeled by the Poisson RFS is located in and also contributes to data cell~$m$.
All possible association vectors $\bar{\assv}_k$ and $\bar{\V{\theta}}_k$ are collected in the sets $\bar{\mathcal{A}}_{k}$ and $\Theta_{\bar{\assv}_k}$, respectively. 
Note that, unlike $\assvR_k$ and $\RV{\theta}_k$, which model the possible contribution of existing objects to data cells, $\bar{\assvR}_k$ and $\bar{\RV{\theta}}_k$ also take into account the possible non-existence of objects and that existing objects are either modeled by one of the Bernoulli components~$j \rmv\in\rmv \Set{J}_{k-1}$ or the Poisson RFS.

\vspace{0mm}
Using $\bar{\assvR}_k$ and $\bar{\RV{\theta}}_k$, we can now rewrite the posterior pdf $f(\RFSst_k|\mev_{1:k})$ in \eqref{eq:UpdEx2} according to 
\vspace{1mm}
\begin{align}
f(\RFSst_k|\mev_{1:k}) &\propto\rmv \sum_{\bar{\assv}_k \in\ist \bar{\mathcal{A}}_{k}} \sum_{\bar{\V{\theta}}_k \in\ist \Theta_{\bar{\assv}_k}} f(\mev_k|\bar{\assv}_k,\rmv\bar{\V{\theta}}_k,\rmv\RFSst_k)\nn \\[-1.5mm]
& \hspace{27mm}\times f(\bar{\assv}_k,\rmv\bar{\V{\theta}}_k,\rmv\RFSst_k|\mev_{1:k-1})\ist.  \label{eq:Post_rw} \\[-4mm]
\nn
\end{align}
Here, $f(\mev_k|\bar{\assv}_k,\rmv\bar{\V{\theta}}_k,\rmv\RFSst_k)$ is given analogously to \eqref{eq:IntensityModel} as 
\vspace{1mm}
\begin{align}
&f(\mev_k|\bar{\assv}_k,\rmv\bar{\V{\theta}}_k,\rmv\RFSst_k) \nn \\[0.5mm] 
&\hspace{5mm}= \Big(\prod_{j \ist\in\ist \Set{J}_{\bar{\V{\theta}}_k}}\rrmv f_1\big(\me_k^{(\bar{\ass}_k^{(j)})}\big|\st_k\big) \Big) \prod_{m \ist\in\ist \mathcal{M} \backslash \mathcal{M}_{\bar{\V{a}}_k}}\rmv f_0\big(\me_k^{(m)}\big)\ist. \label{eq:IntensityModel_2} \\[-6mm]
\nn
\end{align}
The set $\Set{J}_{\bar{\V{\theta}}_k}$ comprises all $j\rmv\in\rmv\Set{J}_k$ with $\theta_k^{(j)} \rmv= 1$.
Furthermore, $f(\bar{\assv}_k,\rmv\bar{\V{\theta}}_k,\rmv\RFSst_k|\mev_{1:k-1})$ is given by (cf. Eq.~\eqref{eq:UpdEx3})
\vspace{1mm} 
\begin{equation}
\label{eq:Post_rw2} 
f(\bar{\assv}_k,\rmv\bar{\V{\theta}}_k,\rmv\RFSst_k|\mev_{1:k-1}) \ist\triangleq\ist p(\bar{\assv}_k,\rmv\bar{\V{\theta}}_k|\RFSst_k)\ist f(\RFSst_k|\mev_{1:k-1}) 
\vspace{1mm}
\end{equation} 
with (cf. Eq.~\eqref{eq:prior3})  
\begin{equation}
\label{eq:Post_rw3} 
p(\bar{\assv}_k,\rmv\bar{\V{\theta}}_k|\RFSst_k) =\ist p(\bar{\V{\theta}}_k|\bar{\assv}_k,\rmv\RFSst_k)\ist \prod_{j\ist\in\ist\Set{J}_{k}} \delta^{(\bar{\ass}_k^{(j)})}(\st_k) 
\vspace{-1mm}
\end{equation}
and
\vspace{1mm}
\begin{equation}
\label{eq:Post_rw4} 
p(\bar{\V{\theta}}_k|\bar{\assv}_k,\rmv\RFSst_k) = \prod_{m \ist\in\ist \Set{M}_{\bar{\assv}_k}}\rmv p^{(m)}\big(\bar{\V{\theta}}^{(m)}_{k}|\RFSst_{k}\big)\ist.
\vspace{-2mm}
\end{equation}
Analogously to $\RV{\theta}^{(m)}_{k}$ as introduced in Section \ref{sec:JLF}, here $\bar{\RV{\theta}}^{(m)}_{k} \rmv= [\rv{\theta}_k^{(m,1)}\rmv\ldots\ist\rv{\theta}_k^{(m,\bar{n}_k^{(m)})}]^{\T}$ with $\bar{n}_k^{(m)}\rmv = |\Set{J}_k^{(m)}|$ comprises all entries $\rv{\theta}_k^{(j)}$ of $\bar{\RV{\theta}}_k$ with $j \rmv\in\rmv \Set{J}_k^{(m)}$ in an arbitrary but fixed order, where $\Set{J}_k^{(m)}$ are all $j$ with $\bar{\ass}_k^{(j)} = m$.
After reformulating $f(\RFSst_k|\mev_{1:k})$ in terms of $\bar{\assvR}_k$ and $\bar{\RV{\theta}}_k$, we are now ready to apply a series of approximations in order to regain a PMB posterior pdf. The first approximation will be described in the next subsection.

\vspace{0mm}
\subsection{Marginalization of $p(\bar{\V{\theta}}_k|\bar{\assv}_k,\rmv\RFSst_k)$}
\label{sec:MargContProb}

\vspace{0mm}
A closer inspection of $f(\bar{\assv}_k,\rmv\bar{\V{\theta}}_k,\rmv\RFSst_k|\mev_{1:k-1})$ shows that it does not further factorize as according to \eqref{eq:Post_rw2}--\eqref{eq:Post_rw4}, which is because of the statistical dependency of the entries $\rv{\theta}_k^{(m,j)}$ of $\RV{\theta}_k^{(m)}$ described by $p(\bar{\V{\theta}}_k|\bar{\assv}_k,\rmv\RFSst_k)$ in \eqref{eq:Post_rw4} (cf. Section \ref{sec:MeaGen1}).
Hence, our first step is to approximate all $\rv{\theta}_k^{(m,j)}$ as independent 
which in turn corresponds to approximate $p(\bar{\V{\theta}}_k|\bar{\assv}_k,\rmv\RFSst_k)$ in \eqref{eq:Post_rw4} by the product of its marginals, i.e.,
\vspace{0.5mm}
\begin{align}
p\big(\bar{\V{\theta}}_k|\bar{\assv}_k,\rmv\RFSst_k\big) &\ist\approx\prod_{j\ist\in\ist\Set{J}_k}  p^{(j,\bar{\ass}_k^{(j)}\rmv\rmv,\theta_k^{(j)})}(\st_k) 
\label{eq:UpdEx5} \\[-6mm]
\nn 
\end{align}
with $p^{(j,\ist\bar{\ass}_k^{(j)}\rmv\rmv,\theta_k^{(j)})}(\st_k) \rmv\triangleq p\big(\theta_k^{(j)}|\bar{\ass}_k^{(j)}\rmv,\st_k\big)$.
Here, $p^{(j,m,1)}(\st_k)$ is the probability that an object modeled by the Bernoulli component $j \rmv\in\rmv \Set{J}_{k-1}$ or the Poisson RFS, i.e., $j \rmv\in\rmv \Set{J}_{k}^{\text{N}}$, contributes to data cell~$m$ and $p^{(j,m,0)}(\st_k)$ that it does not contribute to it.
An exact computation of $p^{(j,\ist\bar{\ass}_k^{(j)}\rmv\rmv,\theta_k^{(j)})}(\st_k)$ is computationally challenging. 

In the following, we present an approximate computation of $p^{(j,\ist\bar{\ass}_k^{(j)}\rmv\rmv,\theta_k^{(j)})}(\st_k)$ for the Swerling 1 object case (cf. Section \ref{sec:MeaGen2}). 
In fact, $p^{(m)}\rmv\big(\V{\theta}^{(m)}_{k} \rmv= \V{\theta}_k^{(m,i)}|\Set{X}_{k} \big)$ is given in the Swerling 1 case by \eqref{eq:priorEtaSwerling}. 
In order to compute $p^{(j,\ist\bar{\ass}_k^{(j)}\rmv\rmv,\theta_k^{(j)})}(\st_k)$ exactly, one needs to marginalize out the potential contributions from all objects except $\st_k$, where $\st_k$ is either modeled by a Bernoulli component or the Poisson RFS.
Note that the complexity of this exact marginalization scales exponentially in the number of objects.
Instead, we propose to replace the exact object contributions $\sigma^{2}_{m}(\st_k)$ of each object $\st_k$ in \eqref{eq:priorEtaSwerling} by an average object contribution, thus avoid an exact marginalization. This leads to
\vspace{0.5mm}
\begin{equation}
\label{eq:upd_MissDet1}
p^{(j,m,1)}(\st_k) \ist\approx\ist \frac{\sigma^{2}_{m}(\st_k)}{\sigma^{2}_{m}(\st_k) + \sum_{j'\in\Set{J}_{k}^{(m)}\setminus \{j\}} \bar{\sigma}_{j,m,k}^{2}}  
\end{equation}
for $j\rmv\in\rmv\Set{J}_k$, $m\rmv\in\rmv\Set{M}$, and with
\begin{align}
&\bar{\sigma}^{2}_{j,m,k} \nn\\[1mm]  
&=\begin{cases}
\ex_{k|k-1}^{(j)}\int \delta^{(m)}(\st_k) \ist\sigma^{2}_{m}(\st_k)\ist f_{k|k-1}^{(j)}(\st_k)\ist \text{d}\st_k, & j\in\Set{J}_{k-1}, \\[2mm]
\int \delta^{(m)}(\st_k) \ist\sigma^{2}_{m}(\st_k)\ist \lambda_{k|k-1}(\st_{k})\ist \text{d}\st_k, & j \in \Set{J}^{\text{N}}_{k}.
\end{cases}
\nn \\
\label{eq:MeanCon1} \\[-6mm]
\nn
\end{align}
Here, $\sigma^{2}_{m}(\st_k) \rmv\triangleq\rmv d^{(m)}(\st_k) \rmv+\rmv \sigma^{2}_{0}$. Recap that $j \rmv=\rmv J_{k-1} + m$ for $j\rmv\in\Set{J}_k^{\text{N}}$.
Note that $\bar{\sigma}^{2}_{j,m,k}$ can be interpreted as the average contribution of an object modeled by the Bernoulli component $j\rmv\in\rmv\Set{J}_{k-1}$ or the Poisson RFS for $j \in \Set{J}_k^{\text{N}}$ to pixel $m \rmv\in\rmv \Set{M}$. 

\vspace{-2mm}
\subsection{PMBM Posterior pdf}
\vspace{0mm}

Next, we insert approximation \eqref{eq:UpdEx5} into \eqref{eq:Post_rw3}, and the resulting expression together with \eqref{eq:IntensityModel_2} and \eqref{eq:Post_rw2} into \eqref{eq:Post_rw}, which yields the approximate posterior pdf displayed in \eqref{eq:AproxPost1}.
\vspace{0mm}
\begin{figure*}[t]
\begin{align} 
f(\RFSst_k|\mev_{1:k})\ist 
&\propto \sum_{\RFSst_{k}^{(1)} \uplus\ist \RFSst_{k}^{(2)} =\ist\RFSst_{k}} \sum_{\bar{\assv}_k \in\ist \bar{\Set{A}}_k} \iist \sum_{\bar{\V{\theta}}_k \in\ist \Theta_{\bar{\assv}_k}} \Big(\prod_{j \in\ist \Set{J}_{k-1}^{(0)}} 1 -\ex_{k|k-1}^{(j)} \Big)  \Big(\prod_{j'\ist\in\ist\Set{J}_{k-1}^{(1)}} \ex_{k|k-1}^{(j)}\ist f\big(\me_k^{(\bar{\ass}_k^{(j')})}\rmv\rmv,\theta_k^{(j')}|\st_k\big) \ist \sd_{k|k-1}^{(j')}(\st_k) \Big)  \nn\\[1mm]
&\times \Big(\prod_{j''\in\ist\Set{J}_k^{\text{N}}} f\big(\me_k^{(\bar{\ass}_k^{(j'')})}\rmv\rmv,\theta_k^{(j'')}|\st_k\big)\ist \lambda_{k|k-1}(\st_k) \Big)\ist \prod_{m \ist\in\ist \mathcal{M} \backslash \mathcal{M}_{\bar{\assv}_k}} f_{0}\big(\me_k^{(m)}\big)  \label{eq:AproxPost1}\tag{39}
\end{align}
\hrule\hrule
\vspace{-3mm}
\setcounter{equation}{39}
\end{figure*}
Here, the set $\Set{J}_{k-1}^{(0)}$ comprises all $j \rmv\in\rmv \Set{J}_{k-1}$ with $\RFSst_k^{(2,j)} \rmv=\rmv \emptyset$ and the set $\Set{J}_{k-1}^{(1)}$ all $j \rmv\in\rmv \Set{J}_{k-1}$ with $\RFSst_k^{(2,j)} \rmv= \{\st_k\}$. 
Furthermore, $f\big(\me_k^{(\bar{\ass}_k^{(j)})}\rmv\rmv,\theta_k^{(j)}|\st_k\big)$ is given for $\bar{\ass}_k^{(j)} = m$ as
\vspace{0mm}
\begin{align*}
&f\big(\me_k^{(m)}\rmv\rmv,\theta_k^{(j)}|\st_k\big) \nn \\[2mm]
&\hspace{4mm}\triangleq
\begin{cases}
\delta^{(m)}(\st_k)\ist p^{(j,m,0)}(\st_k), &\rrmv\rrmv \theta_k^{(j)} = 0\ist, \\[2mm]
\delta^{(m)}(\st_k)\ist p^{(j,m,1)}(\st_k) \ist f_{1}\big(\me_k^{(m)}\big|\st_k\big), &\rrmv\rrmv \theta_k^{(j)} = 1\ist.
\vspace{1.5mm} 
\end{cases}
\nn
\end{align*}

The approximate posterior pdf in \eqref{eq:AproxPost1} is of same form as the posterior pdf as in \cite{Wil:J15}, which, as furthermore shown in \cite{Wil:J15}, can be rewritten into a PMBM pdf.
Thus, we can rewrite \eqref{eq:AproxPost1} as a PMBM pdf (cf. \eqref{eq:fund5}) according to \cite{Wil:J15}
\vspace{1mm}
\begin{equation}
\label{eq:upd1}
f(\RFSst_{k}|\mev_{1:k}) \ist\approx\! \sum_{\RFSst_{k}^{(1)} \uplus\ist \RFSst_{k}^{(2)} =\ist\RFSst_{k}}\hspace{-3mm} f^{\text{P}}\big(\RFSst_{k}^{(1)}\big)\ist f^{\text{MBM}}\big(\RFSst_{k}^{(2)}\big)\ist. 
\vspace{0mm}
\end{equation}
In the following, we provide for $f^{\text{P}}\big(\RFSst_{k}\big)$ and $f^{\text{MBM}}\big(\RFSst_{k}\big)$\vspace{0mm}.

We start with the Poisson pdf $f^{\text{P}}\big(\RFSst_{k}\big)$. In fact, $f^{\text{P}}\big(\RFSst_{k}\big)$ is fully characterized by the posterior PHD $\lambda(\st_k)$ given by
\vspace{1mm}
\begin{equation}
\label{eq:upd_PHD1} 
\lambda(\st_k) = \Big(\sum_{m\ist\in\ist\Set{M}}\delta^{(m)}(\st_k)\ist p^{(0,m,0)}(\st_k)\Big) \ist \lambda_{k|k-1}(\st_{k}) \ist.
\vspace{0.5mm} 
\end{equation}
Here, $p^{(0,m,0)}(\st_k) \triangleq\ist p^{(j,m,0)}(\st_k)$ with $j = J_{k-1} + m$ is the probability that an object modeled by the Poisson RFS does not contribute to data cell $m$ and $\lambda_{k|k-1}(\st_{k})$ is the PHD given by \eqref{eq:pred2}.

Next, we provide an expression for the  MBM pdf $f^{\text{MBM}}\big(\RFSst_{k}\big)$ in \eqref{eq:upd1}. In fact, we have
\begin{equation}
\label{eq:upd3}
f^{\text{MBM}}\big(\RFSst_{k}\big)\ist  =  \sum_{\bar{\assv}_k\ist\in\ist\bar{\mathcal{A}}_{k}} \sum_{\bar{\V{\theta}}_k \in \Theta_{\bar{\assv}_k}} \ist p(\bar{\assv}_k,\rmv\bar{\V{\theta}}_k)\ist\ist f^{\text{MB}}_{\bar{\assv}_k\rmv,\bar{\V{\theta}}_k}(\RFSst_k) 
\end{equation}
where the MB pdfs are given by
\begin{align}
&f^{\text{MB}}_{\bar{\assv}_k,\bar{\V{\theta}}_k}(\RFSst_k) \ist= \rmv \sum_{\RFSst_k^{(1)}\uplus\ldots\uplus\ist\RFSst_k^{(J_k)}=\ist\RFSst_k}\ist \prod_{j \in \Set{J}_k}  \ist f^{(j,\bar{\ass}_k^{(j)}\rmv\rmv,\theta_k^{(j)})}(\RFSst_k^{(j)}) \nn \\[-3mm]
\label{eq:upd3_2} \\[-5mm]
\nn  
\end{align}
and parametrized by the existence probabilities $\ex_k^{(j,\bar{\ass}_k^{(j)}\rmv\rmv,\theta_k^{(j)})}$ and spatial pdfs $\sd^{(j,\bar{\ass}_k^{(j)}\rmv\rmv,\theta_k^{(j)})}(\st_k)$, 
and the association pmf $p(\bar{\assv}_k,\rmv\bar{\V{\theta}}_k)$ by
\vspace{-1mm}
\begin{equation}
\label{eq:upd4}
p(\bar{\assv}_k,\rmv\bar{\V{\theta}}_k) \propto \prod_{j\ist\in\ist\Set{J}_{k}}\ist \assw_k^{(j,\bar{\ass}_k^{(j)}\rmv\rmv,\theta_k^{(j)})}
\vspace{0mm}
\end{equation}
for $\bar{\assv}_k \rmv\in\rmv \bar{\mathcal{A}}_{k}$, $\bar{\V{\theta}}_k \in \Theta_{\bar{\assv}_k}$ and by $p(\bar{\assv}_k,\rmv\bar{\V{\theta}}_k) \rmv= 0$ otherwise.
Note that in the MBM pdf in \eqref{eq:upd3}, each mixture component 
(i) corresponds to exactly one valid pair of association vectors $\bar{\assv}_k$ and $\bar{\V{\theta}}_k$, 
(ii) has $J_k \rmv=\rmv J_{k-1}+M$ Bernoulli components, and 
(iii) is weighted by the probability $p(\bar{\assv}_k,\rmv\bar{\V{\theta}}_k)$.
In the following, we provide expressions for the existence probabilities $\ex_k^{(j,\bar{\ass}_k^{(j)}\rmv\rmv,\theta_k^{(j)})}$\rmv, the spatial pdfs $\sd^{(j,\bar{\ass}_k^{(j)}\rmv\rmv,\theta_k^{(j)})}(\st_k)$, and the association weights $\assw_k^{(j,\bar{\ass}_k^{(j)}\rmv\rmv,\theta_k^{(j)})}$ involved in \eqref{eq:upd3_2} and \eqref{eq:upd4}.

In fact, for $j \rmv\in\rmv\Set{J}_{k-1}$, $\bar{\ass}_k^{(j)}\rmv =\rmv 0$, and $\theta_k^{(j)} \rmv\rmv=\rmv 0$, we have
$\ex_k^{(j,0,0)} \rmv= 0$\ist, $\sd^{(j,0,0)}(\st_k)$ undefined, and 
\vspace{0.5mm}
\begin{align}
\assw_k^{(j,0,0)} &\ist=\ist 1 - \ex_{k|k-1}^{(j)}\ist.  \label{eq:upd_para0_3} \\[-5mm]
\nn
\end{align} 
Here, $\ex_k^{(j,0,0)} \rmv=\rmv 0$ indicates that the object modeled by Bernoulli component $j$ does not exist. Hence, its state pdf $\sd^{(j,0,0)}(\st_k)$ remains undefined and the likelihood of this event is given by \eqref{eq:upd_para0_3}.  

For $j \rmv\in\rmv\Set{J}_{k-1}$, $\bar{\ass}_k^{(j)} = m \rmv\in\rmv \Set{M}$, and $\theta_k^{(j)} = 0$ we have
\vspace{2mm}
\begin{align}
\ex_k^{(j,m,0)} & \ist=\ist 1 \label{eq:upd_para2_2} \\[1.5mm]
\sd^{(j,m,0)}(\st_k) & \ist=\ist \frac{\delta^{(m)}(\st_k)\ist p^{(j,m,0)}(\st_k)\ist \sd_{k|k-1}^{(j)}(\st_k)}{b_k^{(j,m)}} \label{eq:upd_para2_3}   \\[1.5mm]
\assw_k^{(j,m,0)} &\ist=\ist \ex_{k|k-1}^{(j)}\ist b_k^{(j,m)} \label{eq:upd_para2_1}  \\[-2.7mm]
\nn
\end{align}
with $b_k^{(j,m)} \triangleq \int\rmv \delta^{(m)}(\st_k)\ist p^{(j,m,0)}(\st_k)\ist \sd_{k|k-1}^{(j)}(\st_k)\ist \text{d}\st_k$.
Here, $\ex_k^{(j,m,0)} \rmv=\rmv 1$ in \eqref{eq:upd_para2_2} indicates that the object modeled by Bernoulli component $j$ exists, is located in but does not contribute to data cell $m$.
The state of this object is distributed according to \eqref{eq:upd_para2_3} and the likelihood of this event is given by \eqref{eq:upd_para2_1}.

For $j \rmv\in\rmv\Set{J}_{k-1}$, $\bar{\ass}_k^{(j)} = m \rmv\in\rmv \Set{M}$, and $\theta_k^{(j)} = 1$ we have
\vspace{1mm}
\begin{align}
\ex_k^{(j,m,1)} & \ist=\ist 1 \label{eq:upd_para1_2} \\[1.5mm]
\sd^{(j,m,1)}(\st_k) & \ist=\ist \frac{\delta^{(m)}(\st_k)\ist p^{(j,m,1)}(\st_k)\ist f_1\big(\me_k^{(m)}|\st_k\big) \ist \sd_{k|k-1}^{(j)}(\st_k)}{c_k^{(j,m)}} \label{eq:upd_para1_3}  \\[1.5mm]
\assw_k^{(j,m,1)} &\ist=\ist \ex_{k|k-1}^{(j)}\ist c_k^{(j,m)} \label{eq:upd_para1_1}  \\[-2.7mm]
\nn
\end{align} 
with $c_k^{(j,m)} \triangleq \int\rmv \delta^{(m)}(\st_k)\ist p^{(j,m,1)}(\st_k)\ist f_1(\me_k^{(m)}|\st_k) \ist \sd_{k|k-1}^{(j)}(\st_k)$ $\times\text{d}\st_k\ist$.
Here, $\ex_k^{(j,m,1)} \rmv=\rmv 1$ in \eqref{eq:upd_para1_2} indicates that the object modeled by Bernoulli component $j$ exists, is located in and also contributes to data cell $m$. The state of this object is distributed according to \eqref{eq:upd_para1_3} and the likelihood of this event is given by \eqref{eq:upd_para1_1}.

For $j \rmv\in\rmv \Set{J}_k^{\text{N}}$, $\bar{\ass}_k^{(j)} = m$, and $\theta_k^{(j)} = 1$ we have
\vspace{-1.5mm}
\begin{align}
\ex_k^{(j,m,1)} & \ist=\ist \frac{d_k^{(j)}}{f_0(\me_k^{(m)}) + d_k^{(j)}} \label{eq:upd_para3_2} \\[2mm]
\sd^{(j,m,1)}(\st_k) & \ist=\ist \frac{\delta^{(m)}(\st_k)\ist p^{(0,m,1)}(\st_k)\ist f_1\big(\me_k^{(m)}|\st_k\big)\ist \lambda_{k|k-1}(\st_k)}{d_k^{(j)}} \label{eq:upd_para3_3}  \\[2mm]
\assw_k^{(j,m,1)} &\ist=\ist f_0(\me_k^{(m)}) + d_k^{(j)} \label{eq:upd_para3_1}  \\[-4mm]
\nn
\end{align}
with $d_k^{(j)} \triangleq \int \delta^{(m)}(\st_k)\ist p^{(0,m,1)}(\st_k)\ist f_1(\me_k^{(m)}|\st_k)\ist\lambda_{k|k-1}(\st_k)$ $\times\text{d}\st_k$.
Here, \eqref{eq:upd_para3_2} is the probability that an object modeled by the Poisson RFS is located in and contributes to data cell $m$ and that the contribution is not due to clutter. The state of this object is distributed according to \eqref{eq:upd_para3_3} and the likelihood of this event is given by \eqref{eq:upd_para3_1}.  
Finally, for $j \rmv\in\rmv J_{k}^{\text{N}}$, $\bar{\ass}_k^{(j)} = m$, and $\theta_k^{(j)} = 0$, we have $\assw_k^{(j,m,0)} = 1$, $\sd^{(j,m,0)}(\st_k)$ undefined, and $\ex_k^{(j,m,0)}\rmv = 0$.

\vspace{0mm}
\section{Update Step - Stage Two}
\label{sec:up_st_two}
\vspace{.5mm}

After applying the first stage of the update step, the exact posterior pdf in \eqref{eq:UpdEx2} is converted to the PMBM posterior pdf in \eqref{eq:upd1}. 
Next, we apply further approximations that will finally convert the PMBM posterior pdf into a PMB pdf.

\subsection{Marginalization of $p(\bar{\assv}_k,\rmv\bar{\V{\theta}}_k)$}
\label{sec:upd_approx}
\vspace{0.5mm}


We now approximate the MBM pdf $f^{\text{MBM}}\big(\RFSst_{k}\big)$ in \eqref{eq:upd3} by an MB pdf. 
This approximation is in turn based on the approximation of the association pmf $p(\bar{\assv}_k,\rmv\bar{\V{\theta}}_k)$ in \eqref{eq:upd4} by the product of its marginals according to
\vspace{0.5mm}
\begin{equation}
\label{eq:Marg1}
p(\bar{\assv}_k,\bar{\V{\theta}}_k) \ist\approx\ist \prod_{j \in \Set{J}_k} p\big(\bar{\ass}_k^{(j)}\rmv,\theta_k^{(j)}\big) \ist, \quad \bar{\assv}_k \rmv\in\rmv \bar{\mathcal{A}}_{k}, \iist\bar{\V{\theta}}_k \rmv\in\rmv \bar{\Theta}_k
\end{equation}
with $\bar{\Theta}_k \triangleq \{0,1\}^{J_k}$\rmv.
The marginalization is given by
\vspace{1.5mm}
\begin{equation}
\label{eq:Marg2}
p\big(\bar{\ass}_k^{(j)}\rmv,\theta_k^{(j)}\big) =\rmv \sum_{\sim\bar{\assv}_k^{(j)}}\sum_{\sim\bar{\V{\theta}}_k^{(j)}} \ist p(\bar{\assv}_k,\rmv\bar{\V{\theta}}_k) 
\vspace{-1.5mm}
\end{equation}	
where $\sim\rrmv\bar{\assv}_k^{(j)}$ and $\sim \rrmv \bar{\V{\theta}}_k^{(j)}$ denote the vectors of all $\bar{\ass}_k^{(j')}$ and $\theta_k^{(j')}$\rmv, respectively, with $j' \in \Set{J}_k \setminus \{j\}$.
A fast approximate BP-based calculation of \eqref{eq:Marg2} will be presented in Section \ref{sec:ApprBP}.

We proceed by inserting \eqref{eq:Marg1} into \eqref{eq:upd3}, which yields the approximate MBM posterior pdf given\vspace{1mm} by 
\begin{align}
f^{\text{MBM}}(\RFSst_k)  &\ist\approx\rmv \sum_{\RFSst_k^{(1)}\uplus\ldots\uplus\ist\RFSst_k^{(J_k)}=\ist\RFSst_k} \ist \sum_{\bar{\assv}_k\ist\in\ist\bar{\mathcal{A}}_{k}} \ist\sum_{\bar{\V{\theta}}_k\in\bar{\Theta}_k} \nn \\[1mm]
& \hspace{3mm}\times\prod_{j \ist\in\ist \Set{J}_k}\ist\ist p\big(\bar{\ass}_k^{(j)}\rmv,\rmv\theta_k^{(j)}\big) \ist f^{(j,\bar{\ass}_k^{(j)}\rmv\rmv,\theta_k^{(j)})}(\RFSst_k^{(j)}) \ist.\label{eq:approx2} \\[-5mm]
\nn
\end{align} 
As was shown in, e.g., \cite{Kro21TBD}, a pdf of that form
can be rewritten as an MB pdf according to
\begin{align}
f^{\text{MBM}}(\RFSst_k) &\ist\approx\ist f^{\text{MB}}(\RFSst_k)  \nn \\[1mm]
&\ist= \rmv \sum_{\RFSst_k^{(1)}\uplus\ldots\uplus\ist\RFSst_k^{(J_k)}=\ist\RFSst_k} \prod_{j \ist\in\ist \Set{J}_k} \ist f^{(j)}(\RFSst_k^{(j)}) 
\label{eq:approx3_2} \\[-5.5mm]
\nn
\end{align}
where the existence probabilities and spatial pdfs describing the Bernoulli pdfs $f^{(j)}(\RFSst_k)$ are given for $j \rmv\in\rmv \Set{J}_{k-1}$ by
\vspace{0mm}
\begin{align}
\ex_k^{(j)} &=  \sum_{\bar{\ass}_k^{(j)} \in\ist \Set{M}}\ist \sum_{\theta_k^{(j)} =\ist 0}^{1} p(\bar{\ass}_k^{(j)}\rmv\rmv,\rmv\theta_k^{(j)}) \label{eq:approx4} \\[0.5mm]
\sd^{(j)}(\st_k) &= \frac{1}{\ex_k^{(j)}} \rmv\rmv \sum_{\bar{\ass}_k^{(j)} \in\ist \Set{M}}\ist \sum_{\theta_k^{(j)} =\ist 0}^{1} p(\bar{\ass}_k^{(j)}\rmv\rmv,\rmv\theta_k^{(j)})\ist \sd^{(j,\bar{\ass}_k^{(j)}\rmv,\theta_k^{(j)})}(\st_k) \nn \\[-2.5mm]
\label{eq:approx5} \\[-5mm]
\nn
\end{align}
and for $j \rmv\in\rmv \Set{J}_k^{\text{N}}$ by
\vspace{1.5mm}
\begin{align}
\ex_k^{(j)} &=  p(\bar{\ass}_k^{(j)} \rmv=\rmv m\ist,\theta_k^{(j)} = 1)\iist \ex_k^{(j,m,1)} \label{eq:approx6} \\[3mm]
\sd^{(j)}(\st_k) &= \ist \sd^{(j,m,1)}(\st_k)\ist. \label{eq:approx7} \\[-5mm]
\nn
\end{align}
Note that the idea of approximating an MBM pdf by an MB pdf was also used in the derivation  
of the original PMB filter \cite{Wil:J15} and also in, e.g., the BP-based labeled MB filter \cite{Kro19LMB}.

Inserting \eqref{eq:approx3_2} into \eqref{eq:upd1} and thus replacing the MBM pdf $f^{\text{MBM}}(\RFSst_k)$ in \eqref{eq:upd1} with the MB pdf $f^{\text{MB}}(\RFSst_k)$ in \eqref{eq:approx3_2}, yields the approximate posterior pdf given by
\vspace{1mm}
\begin{equation}
\label{eq:ApproxMB}
f(\RFSst_{k}|\mev_{1:k}) \ist\approx\! \sum_{\RFSst_{k}^{(1)} \uplus\ist \RFSst_{k}^{(2)} =\ist\RFSst_{k}}\hspace{-3mm} f^{\text{P}}\big(\RFSst_{k}^{(1)}\big)\ist f^{\text{MB}}\big(\RFSst_{k}^{(2)}\big) \ist.
\end{equation}
Here, the Poisson pdf $f^{\text{P}}(\RFSst_{k})$ is described by the posterior PHD in \eqref{eq:upd_PHD1} and the parameters of the MB pdf $f^{\text{MB}}(\RFSst_{k})$ are given by \eqref{eq:approx3_2}--\eqref{eq:approx7}\vspace{-.5mm}.

\subsection{Recycling of Bernoulli Components}
\label{sec:Recycle}
\vspace{1mm}

The MB pdf $f^{\text{MB}}\big(\RFSst_{k}\big)$ in \eqref{eq:ApproxMB} consists of $J_k = J_{k-1}\rmv+\rmv M$ Bernoulli components, i.e., the number of Bernoulli components increases by $M$ after each time step $k$. 
However, many Bernoulli components typically have a very low existence probability and are thus unlikely to represent an existing object.
In order to reduce computational complexity and memory requirements, we are interested in maintaining a rather small number of Bernoulli components.
Hence, we employ the concept of recycling \cite{Wil12}. Here, all Bernoulli components $j \rmv\in\rmv \mathcal{J}_k^{\text{R}} \rmv\subseteq\rmv \Set{J}_k$ with an existence probability $\ex_k^{(j)}$ below a predefined threshold $\gamma_{\text{R}}$ are approximated by a Poisson RFS by means of moment matching \cite{Wil12}, which results in a single Poisson component for each recycled Bernoulli component.
The overall approximated posterior PHD is given by
\vspace{1mm}
\begin{equation}
\label{eq:approx8}
\tilde{\PHD}(\st_k) = \lambda(\st_k) + \sum_{j\ist\in\ist \mathcal{J}_k^{\text{R}}} \ex_k^{(j)}\ist \sd^{(j)}(\st_k)
\vspace{-1.5mm}
\end{equation}
where $\lambda(\st_k)$ is the posterior PHD given by \eqref{eq:upd_PHD1} and $\ex_k^{(j)}$ and $\sd^{(j)}(\st_k)$ are the Bernoulli parameters given by \eqref{eq:approx4} and \eqref{eq:approx5} or \eqref{eq:approx6} and \eqref{eq:approx7}.
An interpretation of this approximation in terms of the Kullback-Leibler divergence was studied in \cite{Wil12}.  

Summarizing, after applying the exact update step and the approximations described in the previous and this section, the approximate posterior pdf is again of PMB form. In fact, the MB part is represented by the existence probabilities $\ex_k^{(j)}$ and spatial pdfs $\sd^{(j)}(\st_k)$, $j \rmv\in\rmv \Set{J}_k\setminus\mathcal{J}_k^{\text{R}}$ in \eqref{eq:approx4}--\eqref{eq:approx7} and the Poisson part by the PHD $\tilde{\PHD}(\st_k)$ in \eqref{eq:approx8}.
Before we will assess the performance of our proposed algorithm in terms of some simulation experiments, we present a fast approximate computation of the marginal association probabilities $p\big(\bar{\ass}_k^{(j)}\rmv,\rmv\theta_k^{(j)}\big)$ defined in Section \ref{sec:upd_approx}. 


\subsection{Fast BP-Based Computation of $p\big(\bar{\ass}_k^{(j)}\rmv,\theta_k^{(j)}\big)$}
\label{sec:ApprBP}
\vspace{0mm}

The exact computation of $p(\bar{\ass}_k^{(j)}\rmv,\rmv\theta_k^{(j)})$ via \eqref{eq:Marg2} scales exponentially in the number of Bernoulli components and the number of data cells.
In the following, we present a fast BP-based algorithm for an approximate computation of $p(\bar{\ass}_k^{(j)}\rmv,\theta_k^{(j)})$ with only linear complexity. This algorithm is a variant of the algorithm presented in \cite{Wil14}.

Analogously to \cite{Wil14}, we introduce the cell-object association vector $\RV{b}_k \rmv=\! \big[ \rv{b}_k^{(1)} \rmv\cdots\ist \rv{b}_k^{(M)} \big]^{\T}\rmv$ of length $M$. 
The interpretation of the entries $\rv{b}_k^{(m)} \!\rmv\in\! \{0\} \cup \Set{J}_{k-1}$ with $m \!\in\! \Set{M}$ is as follows: $\rv{b}_k^{(m)} \!\rmv=\rmv 0$ indicates that the contribution to cell $m$ is either due to an object modeled by the Poisson RFS or due to clutter; 
$\rv{b}_k^{(m)} = j \rmv\in  \Set{J}_{k-1}$ indicates that the contribution to cell $m$ is due to the object modeled by Bernoulli component~$j$.
Using $\RV{b}_k$, we find (cf. Eq. \eqref{eq:upd4})
\begin{equation}
\label{eq:JoiAssPMF}
p(\bar{\assv}_k,\bar{\V{\theta}}_k,\V{b}_k) \propto\ist \Psi(\bar{\assv}_k,\bar{\V{\theta}}_k,\V{b}_k) \prod_{j \ist\in\ist \Set{J}_k}\ist \assw_k^{(j,\bar{\ass}_k^{(j)}\rmv,\theta_k^{(j)})}.
\end{equation}
Here, the admissibility constraint $\Psi(\bar{\assv}_k,\bar{\V{\theta}}_k,\V{b}_k)$ is given by
\vspace{1mm}
\begin{equation}
\label{eq:AdmCon}
\Psi(\bar{\assv}_k,\bar{\V{\theta}}_k,\V{b}_k) =\rmv \prod_{j \ist\in\ist \Set{J}_{k-1}} \ist \prod_{m\ist \in\ist \Set{M}}  \psi_{j,m}(\bar{\ass}_{k}^{(j)}\rmv,\rmv\theta_k^{(j)}\rmv,\rmv b_{k}^{(m)})
\end{equation}
where $\psi_{j,m}(\bar{\ass}_{k}^{(j)}\rmv,\theta_k^{(j)}\rmv, b_{k}^{(m)}) \rmv=\rmv 0$ if either $\bar{\ass}_{k}^{(j)}= m$\ist, $\theta_k^{(j)} = 1$, and $b_{k}^{(m)} \!\rmv\neq\! j$ or $\bar{\ass}_{k}^{(j)} \!\neq\!\rmv m$ and/or $\theta_k^{(j)} = 0$ and $b_{k}^{(m)} \!\rmv=\! j$, and $\psi_{j,m}(\bar{\ass}_{k}^{(j)}\rmv,\theta_k^{(j)}\rmv, b_{k}^{(m)}) \rmv=\! 1$ otherwise. 

\begin{figure}
\hspace*{10mm}
\includegraphics[scale=1]{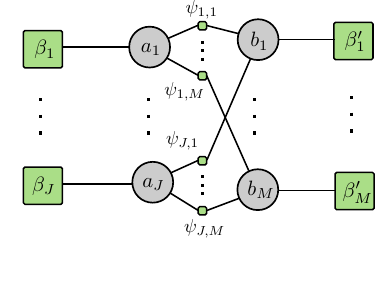}
\vspace{-9mm}
  \caption{Factor graph representation of $p(\bar{\assv}_k,\bar{\V{\theta}}_k,\V{b}_k)$ in \eqref{eq:JoiAssPMF}, \eqref{eq:AdmCon}. 
Variable nodes are depicted as circles and factor nodes as squares. We use the shorthands $J \rmv\triangleq\rmv J_{k-1}$, 
\vspace{-0.3mm}
$\ass_j \triangleq [\bar{\ass}_{k}^{(j)}\theta_k^{(j)}]^{\T}$\rmv, $b_m \triangleq b_k^{(m)}$, 
\vspace{0.3mm}
$\assw_j \rmv\triangleq \assw_k^{(j,m,\theta_k^{(j)})}$\rmv, 
$\beta'_m \triangleq\assw_k^{(j,m,1)}$ with $j = J_{k-1} + m$, and $\psi_{j,m} \rmv\triangleq \psi_{j,m}(\bar{\ass}_{k}^{(j)}\rmv,\theta_k^{(j)}\rmv,b_{k}^{(m)})$.}
	\label{fig:fact}
\vspace{-3mm}
\end{figure}

Based on \eqref{eq:JoiAssPMF}, we can devise an efficient algorithm for calculating accurate approximations to the marginal association pmfs $p(\bar{\ass}_k^{(j)}\rmv,\theta_k^{(j)})$. 
This algorithm is based on applying the BP algorithm, also known as sum-product algorithm \cite{Ksh01}, on the factor graph displayed in Fig.\,\ref{fig:fact}. 
In BP iteration $p \rmv\in\rmv \{1,\ldots,P\}$, a message $\zeta^{[p]}_{j\rightarrow m}$ is passed from variable node ``$[\bar{\ass}_{k}^{(j)}\ist \theta_k^{(j)}]^{\T}\rmv$'' via factor node ``$\psi_{j,m}(\bar{\ass}_{k}^{(j)}\rmv,\theta_k^{(j)}\rmv,b_{k}^{(m)})$'' to variable node ``$b_{k}^{(m)}\rmv$'', and a message $\nu^{[p]}_{m \rightarrow j}$ is passed from variable node ``$b_{k}^{(m)}\rmv$'' via factor node ``$\psi_{j,m}(\bar{\ass}_k^{(j)}\rmv,\theta_k^{(j)}\rmv, b_{k}^{(m)})$'' to variable node ``$[\bar{\ass}_{k}^{(j)}\ist\theta_k^{(j)}]^{\T}\rmv$''. Similarly to \cite{Wil14}, one obtains 
\begin{equation}
\label{eq:BP1}
\zeta^{[p]}_{j\rightarrow m} = \frac{\assw_k^{(j,m,1)}}{\beta_k^{(j,0)} + \sum_{m' \in\ist \Set{M} \setminus \{m\}} \assw_k^{(j,m',1)} \ist \nu^{[p-1]}_{m'\rightarrow j}} 
\vspace{1mm}
\end{equation}
for $j \!\in\! \Set{J}_{k-1}$ and with $\beta_k^{(j,0)} \triangleq \assw_k^{(j,0,0)} + \sum_{m\ist\in\ist\Set{M}}\assw_k^{(j,m,0)}$; and
\vspace{1mm}
\begin{equation}
\label{eq:BP2} 
\nu^{[p]}_{m \rightarrow j} = \frac{1}{\beta_k^{\prime(m,0)} + \sum_{ j' \in\ist \Set{J}_{k-1} \rmv\setminus \{j\}}
  \zeta^{[p]}_{j'\rightarrow m}} 
\end{equation}
for $m \!\in\! \mathcal{M}$ and with $\beta_k^{\prime(m,0)} \triangleq \assw_k^{(j,m,1)}$ where $j = J_{k-1} + m$. 
The recursion established by \eqref{eq:BP1} and \eqref{eq:BP2} is initialized 
by $\nu^{[0]}_{m\rightarrow j} \!=\! 1$. After the final iteration $p \rmv=\! P$, beliefs $\tilde{p}(\bar{\ass}_k^{(j)}\rmv,\theta_k^{(j)})$ for the respective variable nodes ``$[\bar{\ass}_k^{(j)}\ist\theta_k^{(j)}]^{\T}$'' are calculated for $j\in\Set{J}_{k-1}$ according to  
\vspace{0mm}
\be
\label{eq:BPmarg3}
\tilde{p}(\bar{\ass}_{k}^{(j)} \rmv= m,\theta_k^{(j)}) =
\begin{cases}
 \assw_k^{(j,m,1)}\ist \nu^{[P]}_{m \rightarrow j}/N_k^{(j)}, & \theta_k^{(j)}  = 1 \\[2mm]
 \assw_k^{(j,m,0)}/N_k^{(j)}, & \theta_k^{(j)} = 0
\end{cases}
\vspace{.5mm}
\ee
with $N_k^{(j)}\triangleq \assw_k^{(j,0)} \rmv+ \sum_{m' \in \Set{M}} \assw_k^{(j,m',1)} \ist \nu^{[P]}_{m' \rightarrow j}$, 
\vspace{0.5mm}
and for $j = J_{k-1} + m$ with $m\in\Set{M}$ (which is equal to $j\in\Set{J}_k^{\text{N}}$) according to
\be
\label{eq:BPmarg4}
\tilde{p}(\bar{\ass}_{k}^{(j)} \rmv= m,1) =  \frac{1}{\beta_k^{\prime(m,0)} + \sum_{j \in\ist \Set{J}_{k-1}} \zeta^{[P]}_{j\rightarrow m}}.
\vspace{.5mm}
\ee
The beliefs $\tilde{p}(\bar{\ass}_k^{(j)}\rmv,\theta_k^{(j)})$ provide approximations to the marginal association pmfs $p(\bar{\ass}_k^{(j)}\rmv,\theta_k^{(j)})$, $j \!\in\! \mathcal{J}_k$\vspace{1mm}.

\section{Simulation Study}
\label{sec:num}

\vspace{1mm}
\subsection{Simulation Setup and Simulation Methods}
\vspace{0.5mm}

\begin{figure}[t!]
 \vspace*{1mm}
\centering
\hspace{5mm}
\footnotesize
{\scalebox{1}{\hspace{-2mm}\includegraphics[scale=.8]{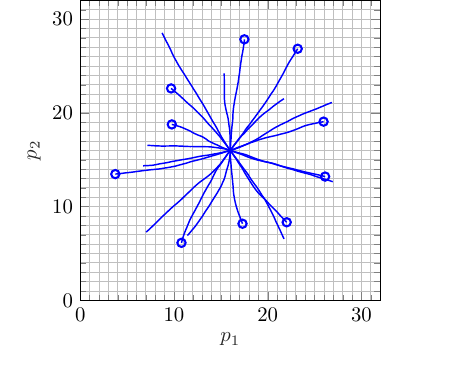}}}
\vspace{-3mm}
\caption{Object trajectories; the starting positions are indicated by blue circles\vspace{-2mm}.}
\label{fig:Traj} 
\vspace{-4mm}
\end{figure}	

We consider a 2D simulation scenario with a region of interest (ROI) of $[\text{0}\text{m},\text{32}\text{m}] \times [\text{0}\text{m},\text{32}\text{m}]$. 
We simulated $10$ objects during $200$ time steps. The objects' trajectories are visualized in Fig.\! \ref{fig:Traj}; the objects appear at various times before time step $30$, they come close to each other around time step $100$, then they separate again and disappear at various times after time step $170$.
We consider two different scenarios, denoted as S1 and S2. 
In S1, all objects have an intensity of $\gamma_{\text{I}} \rmv=\rmv 10$, and in S2 an intensity of $\gamma_{\text{I}} \rmv=\rmv 4$.

The object states consist of 2D position, 2D velocity, and the object's intensity (cf.\! Section\! \ref{sec:sys}), i.e., 
$\stR_k \!=\rmv [\RV{p}_{k}^{\T} \,\, \rv{\gamma}_k]^{\T}\rmv$ with $\RV{p}_{k} \rmv=\rmv [\rv{p}_{k,1} \,\, \rv{p}_{k,2} \,\, \dot{\rv{p}}_{k,1} \,\, \dot{\rv{p}}_{k,2}]^{\T}$\rmv. 
In the tracking methods, the state transition is modeled as follows. 
The kinematic part $\RV{p}_{k}$ evolves according to a nearly constant velocity motion model, which is given by $\RV{p}_k \rmv= \M{A}\ist \RV{p}_{k-1} + \M{W}\ist \RV{w}_k$ with $\M{A} \rmv\in\rmv \mathbb{R}^{4\times 4}$ and $\M{W} \rmv\in\rmv \mathbb{R}^{4\times 2}$ chosen as in \cite{Kro16}. The driving noise $\RV{w}_k$ is distributed according to $\Set{N}(\V{w}_{k};\V{0}_2,10^{-3}\M{I}_2)$. 
The object's intensity evolves according to a random walk model, which is given by $\rv{\gamma}_k \rmv= \rv{\gamma}_{k-1} + \rv{\epsilon}_k$.
The driving noise $\rv{\epsilon}_k$ is distributed as $\mathcal{N}(\epsilon_{k};0,10^{-2})$.

The measurement is an image consisting of $32 \times 32$ cells covering the ROI.
Each cell is a square of $1\text{m}$ side length and has a scalar intensity.
The cell intensities are modeled according to the Swerling 1 data model introduced in Section \ref{sec:MeaGen2}.
More precisely, the cell intensities are generated according to expression \eqref{eq:Meas1} and setting 
\[
d^{(m)}(\st_k^{(i)}) =
\begin{cases}
\gamma_k^{(i)}, & \text{object } \st_k^{(i)} \text{ is in cell } m \\[1mm]
0, & \text{otherwise.}
\end{cases}
\vspace{0mm}
\]
This leads to Rayleigh distributed cell intensities given by $\Set{R}\big(\me^{(m)}_k; \sum_{i \ist\in\ist \Set{I}_{k}^{(m)}} \sqrt{\gamma^{(i)}_k} + \sigma_{\text{n}}\big)$, where the set $\Set{I}_{k}^{(m)}$ comprises all indexes of objects that are located in cell $m$ (cf. Section~\ref{sec:MeaGen1}). 
If there is no object located in data cell $m$, the cell intensity is distributed according to $\Set{R}\big(\me^{(m)}_k; \sigma_{\text{n}}\big)$.
The noise variance $\sigma_{\text{n}} = 1$ which leads to an SNR of $10\text{dB}$ for S1 and $6\text{dB}$ for S2.

We employ a particle implementation of our proposed association-based PMB filter with object contribution probabilities, briefly referred to as PMBF-AC.
We compare PMBF-AC with particle implementations of the association-based PMB filter in \cite{Kro21TBD}, referred to as PMBF-A, the multi-Bernoulli filter in \cite{Vo10TBD}, simply referred to as MBF and the MB filter based on a superpositional measurement model \cite{Lia23TBD,Dav24TBD}, referred to as MBF-S. As mentioned in Section~\ref{sec:contr}, PMBF-A is not based on any random object contribution model. 

PMBF-AC, PMBF-A, and MBF represent the spatial pdf of each Bernoulli component by $3,\!000$ particles while MBF-S represent the spatial pdf of each Bernoulli component by only $100$ particles. Even with only $100$ particles, MBF-S yields a significantly higher runtime compared to the other methods and a further increase of the number of particles in MBF-S did not lead to a notably improved performance.

PMBF-AC and PMBF-A represent the posterior PHD by $50,\!000$ particles and the birth PHD by another $50,\!000$; the resulting $100,\!000$ particles are reduced to $50,\!000$ again after the update step.
More precisely, the birth PHD is given by $\PHD_{\text{B}}(\st_k) \rmv=\rmv \mu_{\text{B}}\ist f_{\text{B}}(\st_k)$ with $\mu_{\text{B}} \rmv=\rmv 0.1$ and $f_{\text{B}}(\st_k) \rmv= f(p_{k,1}, p_{k,2})\ist f_{\text{v}}\big(\dot{p}_{k,1},\dot{p}_{k,2}\big)\ist f_{\text{I}}(\gamma_{k})$ where $f(p_{k,1},p_{k,2})$ is uniform over the ROI, $f_{\text{v}}\big(\dot{p}_{k,1},\dot{p}_{k,2}\big)$ is $\mathcal{N}(\dot{p}_{k,1},\dot{p}_{k,2}; \V{0}_2, \sigma^2_{\text{v}}\ist\textbf{I}_2 )$ with $\sigma^2_{\text{v}} = 0.1$, and $f_{\text{I}}(\gamma_{k})$ is uniform from $0$ to $30$.
MBF and MBF-S generate a new Bernoulli component for each cell measurement $\me_{k-1}^{(m)}$ whose intensity value is above the threshold $\eta_{\text{new}} \rmv=\rmv \sqrt{\gamma_{\text{I}} + \sigma^2_{\text{n}}}$. 
In fact, the existence probabilities of a new Bernoulli component in MBF and MBF-S are set to $10^{-3}$ and the spatial pdfs are  represented by drawing particles from the pdf $f'_{\text{B}}(\st_k) \propto  \int\rmv f( \st_{k} | \st_{k-1}) \ist f\big(\me^{(m)}_{k-1}\ist\big|\ist p_{k-1,1}, p_{k-1,2}\big)$ $\times f_{\text{v}}\big(\dot{p}_{k-1,1},\dot{p}_{k-1,2}\big) f_{\text{I}}(\gamma_{k-1}) \ist\text{d}\st_{k-1}$. 
The function $f\big(\me_{k-1}^{(m)}\big|$ $p_{k-1,1}, p_{k-1,2}\big)$ is uniform over the cell area of measurement~$m$ and $f_{\text{v}}\big(\dot{p}_{k-1,1},\dot{p}_{k-1,2}\big)$ and $f_{\text{I}}(\gamma_{k-1})$ are given as in $f_{\text{B}}(\st_k)$.
We set the survival probability to $\su \rmv=\rmv 0.999$ and the recycling threshold (cf. Section~\ref{sec:Recycle}) in PMBF-AC and PMBF-A to $\eta_{\text{R}} \rmv=\rmv 0.1$\vspace{0mm}.

\begin{figure}[t!]
\footnotesize
\centering
{\hspace*{5mm}\scalebox{1}{\hspace{-2mm}\includegraphics[scale=0.6]{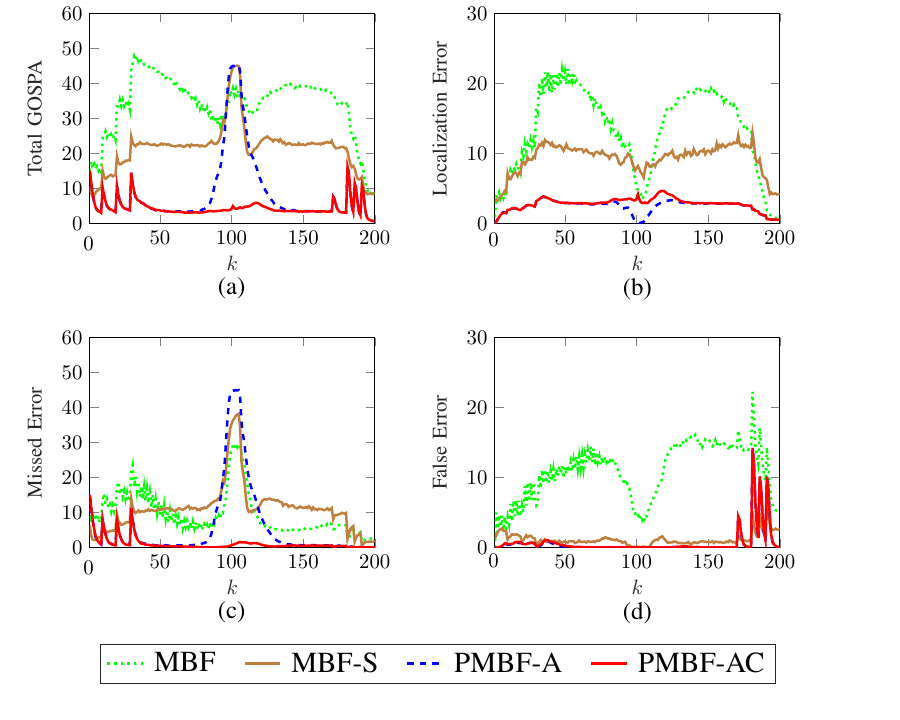}}}
\vspace{-5mm}
\caption{Total GOSPA error (a) and its components (b)--(d) of MBF, MBF-S, PMBF-A, and PMBF-AC versus time $k$ for $\gamma_{\text{I}} = 10$, i.e., an SNR of $10\text{dB}$.}
\label{fig:10dB} 
\vspace{-4mm}
\end{figure}	

\begin{figure}[t!]
\vspace{3mm}
\footnotesize
\centering
{\hspace*{5mm}\scalebox{1}{\hspace{-2mm}\includegraphics[scale=0.6]{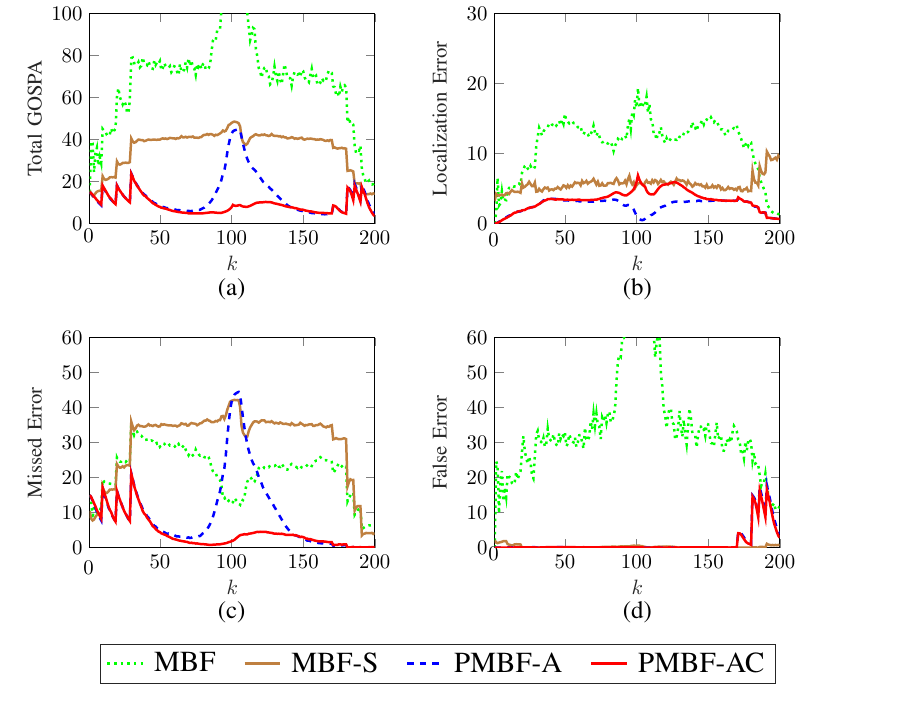}}}
\vspace{-5mm}
\caption{Total GOSPA error (a) and its components (b)--(d) of MBF, MBF-S, PMBF-A, and PMBF-AC versus time $k$ for $\gamma_{\text{I}} = 4$, i.e., an SNR of $6\text{dB}$.}
\label{fig:3dB} 
\vspace{-4mm}
\end{figure}

\subsection{Simulation Results}
\vspace{0mm}

For a quantitative assessment and comparison of the various filters, we employ the Euclidean distance-based generalized optimal subpattern assignment (GOSPA) error metric \cite{Rah17GOSPA} with parameters $c \rmv=\rmv 10$, $p \rmv=\! 1$, and $\alpha = 2$, averaged over 500 simulation Monte Carlo runs. 
The GOSPA error can be decomposed into a localization error, an error due to missed objects (``missed error''), and an error due to false objects (``false error'').
Fig.\! \ref{fig:10dB} and Fig.\! \ref{fig:3dB} show the obtained results of PMBF-AC, PMBF-A, MBF-S, and MBF for $\gamma_{\text{I}} \rmv=\rmv 10$ and for $\gamma_{\text{I}} \rmv=\rmv 2$, respectively. 
The figures indicate that the proposed PMBF-AC outperforms PMBF-A, MBF-S, and MBF. 
More precisely, PMBF-AC and PMBF-A outperform MBF-S and MBF throughout all simulation times, thus indicating a very good performance of association-based approaches. 
When objects come close to each other around time step $k = 100$, PMBF-AC is able to resolve such a situation better than PMBF-A by employing our random object contribution model, resulting in a much lower GOSPA error of PMBF-AC than PMBF-A around time step $k = 100$. 
More precisely, since PMBF-A does not model the interaction of objects, the missing error component (cf. Fig. \ref{fig:10dB}(c) and \ref{fig:3dB}(c)) drastically increases when objects interact representing the incorrect pruning of Bernoulli components that model existing objects\vspace{-1.5mm}.

\section{Conclusion}
\vspace{-.5mm}
\label{sec:con}

We proposed a track-before-detect (TBD) multiobject tracking method based on a new measurement model relying on probabilistic object-to-cell contributions.
Based on the new model, we derived a Poisson multi-Bernoulli (PMB) filter for TBD that uses scalable, belief propagation (BP) based probabilistic data association to achieve moderate runtimes despite the potentially large number of data cells used as measurements. 
Our numerical results demonstrated that the proposed PMB filter for TBD can achieve very high tracking accuracy and outperforms existing TBD methods in scenarios with multiple interacting objects, i.e., in scenarios where multiple objects come close to each other and potentially contribute to the same resolution cells.
A possible direction of future research is the extension to scenarios where objects contribute to multiple data cells by adopting BP for mulitple-measurement to object associations \cite{Mey21EOT} or an extension to neural-enhanced processing \cite{LiaMey:J24}\vspace{0mm}.

\vspace{0mm}
\appendices
\renewcommand*\thesubsectiondis{\thesection.\arabic{subsection}}
\renewcommand*\thesubsection{\thesection.\arabic{subsection}}

\section{} 
\label{sec:App_Limit}
\vspace{-1mm}

In this appendix, we derive expression \eqref{eq:priorEtaSwerling}, i.e., the object contribution prior pmf $p^{(m)}\big(\V{\theta}^{(m)}_{k}|\Set{X}_{k} \big)$ for multiple Swerling~1 objects. 
The procedure is as follows: we first compute $p^{(m)}_{\eta}\big(\V{\theta}^{(m)}_{k}|\Set{X}_{k} \big)$ by inserting $p_{\eta}^{(i,m)}$ in \eqref{eq:Det3} into \eqref{eq:priorEta2} and then 
take the limit $\eta\rmv\to\rmv 0$ according to \eqref{eq:probDet}.
Unfortunately, the resulting expression of $p^{(m)}_{\eta}\big(\V{\theta}^{(m)}_{k}|\Set{X}_{k} \big)$ is  a 'zero-by-zero' division.
However, such limits can often still be computed by applying the rule of de L'Hospital \cite{Kra04}, which states that for two  differentiable functions $f(\eta)$, $g(\eta)$, with $\lim_{\eta \ist\to\ist c} f(\eta) = \lim_{\eta \ist\to\ist c} g(\eta) = 0$, and $g'(\eta)\rmv\neq\rmv 0$, we have
\vspace{-.5mm}
\begin{equation}
\label{eq:lim1}
\lim_{\eta\ist\to\ist c} \frac{f(\eta)}{g(\eta)} \ist=\ist \lim_{\eta\ist\to\ist c} \frac{f'(\eta)}{g'(\eta)}
\vspace{0.5mm}
\end{equation}
where $f'(\eta)$ and $g'(\eta)$ are the first derivatives of $f(\eta)$ and $g(\eta)$, respectively, with respect to $\eta$.
Thus, the limit in $c$ for the quotient of $f(\eta)$ and $g(\eta)$ can be computed by taking the limit of the quotient of $f'(\eta)$ and $g'(\eta)$. 
If $\lim_{\eta\ist\to\ist c} \frac{f'(\eta)}{g'(\eta)}$ is another 'zero-by-zero' division, the attempt can be made to apply the rule of de L'Hospital sequentially until a finite limit is found.

We now take the first derivative of the numerator and denominator of $p^{(m)}_{\eta}\big(\V{\theta}^{(m)}_{k}|\Set{X}_{k} \big)$ in \eqref{eq:priorEta2} with respect to $\eta$.   
Since the resulting limit is again a `zero-by-zero' division, we repeat this procedure 
$n_k^{(m)}\rmv-1$ times, which leads to
\[
p^{(m)}\big(\V{\theta}^{(m)}_{k} \rmv= \V{\theta}_k^{(m,i)}|\Set{X}_{k} \big) = \frac{\frac{1}{\prod_{i'\rmv\in\Set{I}^{(m)}_k \setminus \{i\}}\sigma^2_{\rmv m,i'}}}{\sum_{j\in\ist\Set{I}^{(m)}_k } \frac{1}{\prod_{j'\rmv\in\Set{I}^{(m)}_k \rmv\setminus\rmv \{j\}}\sigma^2_{\rmv m,j'}}} 
\]
where the vector $\V{\theta}_k^{(m,i)}$ is defined as the $i$'s entry of $\V{\theta}_k^{(m)}$ being one and all the other entries being zero (cf. Section \ref{sec:MeaGen1}).
By formally multiplying and dividing by $\prod_{i\ist\in\ist\Set{I}^{(m)}_k } \sigma_{m,i}^2$, we get\vspace{0mm}
\begin{align*}
p^{(m)}\big(\V{\theta}^{(m)}_{k} \rmv= \V{\theta}_k^{(m,i)}|\Set{X}_{k} \big) &= \frac{\sigma^{2}_m(\st_k^{(i)})}{\sum_{i'=1}^{n^{(m)}_k}\sigma^{2}_m(\st_k^{(i')})} \nn \\[-5mm]
\end{align*}
which is equal to \eqref{eq:priorEtaSwerling}.

\renewcommand{\baselinestretch}{.98}
\selectfont
\bibliographystyle{IEEEtran}
\bibliography{references_Kr}

\end{document}